\documentclass[pra,aps,twocolumn,floatfix,superscriptaddress]{revtex4-2}
\usepackage{amsmath}
\usepackage{amssymb}
\usepackage{graphicx}
\usepackage{color}
\usepackage{amssymb}
\usepackage{soul}
\usepackage[utf8]{inputenc}
\usepackage[T1]{fontenc}
\usepackage{array}
\usepackage{ulem}
\begin{document}
\title{Dynamics of quantum soliton in Lee-Huang-Yang spin-orbit coupled Bose-Einstein condensates}
\author{Sonali Gangwar}
\affiliation{Department of Physics, Indian Institute of Technology, Guwahati 781039, Assam, India} 
\author{Rajamanickam Ravisankar}
\affiliation{Department of Physics, Indian Institute of Technology, Guwahati 781039, Assam, India} 
\affiliation{Institute of Atomic and Molecular Sciences, Academia Sinica, Taipei 10617, Taiwan}
\author{Paulsamy Muruganandam}
\affiliation{Department of Physics, Bharathidasan University, Tiruchirappalli 620024, Tamilnadu, India}
\author{Pankaj Kumar Mishra}
\affiliation{Department of Physics, Indian Institute of Technology, Guwahati 781039, Assam, India}

\date{\today}

\begin{abstract} 
We present the numerical results of the structure and dynamics of the self-bound ground state arising solely because of the presence of beyond mean field quantum fluctuation in spin-orbit coupled binary Bose-Einstein condensates in one dimension. Depending upon spin-obit (SO) and Rabi couplings, we observe that the ground state exhibits either quantum-bright (plane) or quantum-stripe soliton nature. We find an analytical soliton solution for non-zero SO coupling that matches quite well with the numerical results. Further, we investigate the dynamical stability of these solitons by adopting three protocols, such as (i) adding initial velocity to each component, (ii) quenching the SO and Rabi coupling parameters at initial and finite time, and (iii) allowing collision between the two spin-components by giving equal and opposite direction velocity to them. Many interesting dynamical features of the solitons, like, multi-fragmented, repelling, and breathing in time and space-time, are observed. For given Rabi coupling frequency, the breathing frequency of the soliton increases upon the increase in SO coupling, attaining a maximum at the critical SO coupling where the phase transition from the bright to stripe soliton occurs. We observe that the maximum breathing frequency exhibits power law dependence on the Rabi coupling frequency with an exponent $\sim 0.16$. The quenching of SO and Rabi coupling frequency leads to dynamical phase transition from bright to stripe soliton and from breathing stripe to a bright soliton. In the absence of the SO and Rabi couplings, depending upon the velocity of the up-spin and down-spin components, the collision between them is either elastic or inelastic, which is consistent with the earlier numerical and experimental observation. However, in the presence of coupling parameters, the collision appears to be inelastic and quasi-elastic in nature.
\end{abstract}

\flushbottom

\maketitle
\section{Introduction}
An exciting aspect of Bose-Einstein condensates (BECs) is the formation of the quantum droplet, which appears in the form of small clusters of atoms bound together owing to the balance between the attractive and repulsive forces. In a binary mixture of the condensate, the quantum droplet manifests in a spherical shape due to the competition between the effective short-range attractive interaction between the atoms and the repulsive interaction arising solely due to quantum fluctuation, which is also responsible for its stabilization~\cite{Luo2021}. Theoretically, one can model the contribution of the quantum fluctuations by considering the beyond mean-field (BMF) term, popularly known as the Lee-Huang-Yang (LHY) correction in the mean-field Hamiltonian of the condensate~\cite{Lee1957}. BMF terms were corroborated first time in experiments with homogeneous and single component Bose gas of $^{85}$Rb~\cite{Papp2008} and $^{7}$Li~\cite{Navon2011}, and recently in binary $^{39}$K~\cite{Skov2021}. In general, the LHY correction term is attractive in quasi-one dimension, while it is repulsive in two and three dimension~\cite{Luo2021, Petrov2015}. 

The advancement in the state-of-art technology and simulation tools in ultracold BECs led to an upsurge in the research in quantum droplets during the past few decades. The quantum droplet was first reported experimentally in the dipolar BECs, for which a single component-dipolar BECs made of dysprosium ($^{164}$Dy) produced elongated quantum droplets in one direction~\cite{Barbut2016, Schmitt2016, Barbut2016b}. Chomaz et al. observed similar features with the condensate of erbium atoms ($^{166}$Er)~\cite{Chomaz2016}. Subsequently, several groups realized quantum droplets in binary mixtures of isotropic BECs~\cite{Cabrera2018, Cheiny2018, Semeghini2018, Ferioli2019},  Bose-Fermi mixture~\cite{Rakshit:2019} and also in binary magnetic gases~\cite{Smith:2021}. These quantum droplets are commonly observed in three dimensions, while it is possible to increase the lifetime of the droplets in the lower dimensions. The formation of one-dimensional quantum droplets is due to the balance between the repulsive mean-field (MF) contribution to the energy per particle, which is linear in the density ($n$) of the gas, and the attractive BMF correction, proportional to $-n^{1/2}$~\cite{Petrov2015, Petrov2016}.  

Since the proposition of the non-trivial attractive nature of the LHY term in quasi-one dimensional quantum droplet~\cite{Petrov2016}, it has caught the great attention of the scientific community. A significant number of experimental~\cite{Frolian2022}, as well as theoretical and numerical works, have been performed in recent years using the effective one component~\cite{Astrakharchik2018}, binary~\cite{Mistakidis2021} and Spin-orbit (SO) and Rabi coupled binary BECs~\cite{Tononi2019, Sahu2020} that explored the structure and dynamics of the Quantum droplet. Tononi et al., in SO-coupled BECs, demonstrated that the self-bound states are solitonic in nature for vanishing mean-field contribution~\cite{Tononi2019}. Depending upon the Rabi and SO coupling parameters range, these soliton-like states are either of single peak (bright soliton) or multiple peaks (stripe soliton) nature~\cite{Chiquillo2018, Tononi2019}. Some of these works demonstrate the existence of soliton and droplet nature of the self-bound state for the effective-one component binary BECs in one dimension ~\cite{Pathak:2022} as well as in the binary mixture with SO coupling in two dimension~\cite{Li2017, Sahu2020}. The transition between the quantum soliton and quantum droplet regime depends on several parameters, like, the atom number ($N$), interaction strength, the strength of the confining potential, etc.~\cite{Pathak:2022}. For large $N$, a highly dense droplet solution exists, while low dense bright soliton occurs for small $N$~\cite{Astrakharchik2018}. A recent experiment shows that the attractive mixture of BECs confined in an optical waveguide could exhibit both quantum soliton and quantum droplets. At large $N$, the droplet exhibits similar characteristics as those of the classical droplets~\cite{Cheiney2018}.  

Though lots of emphasis on exploring the structure and stability of quantum droplets in the recent past, only a limited number of works are available in the literature that focuses on the dynamical aspect of quantum droplets. Ferioli et al. analyzed the dynamics of quantum droplets in a binary mixture by allowing the droplets to collide with each other~\cite{Ferioli2019}. In classical droplet collisions, there are two possibilities, either they merge into a single one, or the colliding droplets separate two or more ones after collision~\cite{Pan2008}. Ferioli et al. have investigated the collision between the quantum droplets by experimental and numerical means for the binary mixture of hyperfine states in $^{39}$K and found the merging and separation between the droplets depending on their velocity. Critical velocity for the transition between the elastic and inelastic collision strongly depends upon the atom number $N$~\cite{Ferioli2019}. A few theoretical studies are available that establish the robustness of the quantum soliton or quantum droplet during the collision. For instance, Astrakharchik and Malomed numerically investigated the static and dynamical properties of quantum droplets using the mean-field theory and studied the collision properties of two counter-propagating droplets~\cite{Astrakharchik2018}. They showed that collisions between the tiny droplets are quasi-elastic, indicating solitonic behaviour. While depending on the velocity, the large drops may undergo merging or fragmentation, which shows the collision as inelastic. Recently, Young and Adhikari numerically studied the collision properties of bright solitons in two-dimensional dipolar BECs~\cite{Young2021}.

In recent years, one may witness numerous theoretical and numerical works focusing on the structure and dynamics of the droplets in binary BECs. However, only a few studies are available on the droplets in the spinor BECs, especially in spin-orbit coupled binary BECs~\cite{Tononi2019, Lin2011}. In particular, the role of SO and Rabi coupling parameters in dictating the shape and dynamics of the quantum droplet is not well understood. In this paper, we present a detailed numerical investigation to understand the effect of the SO and Rabi coupling parameters on the stability and shape of the quantum soliton. Although Tononi \textit{et al.}~\cite{Tononi2019} demonstrates the presence of the self-bound quantum soliton due to the LHY term with vanishing mean-field contributions, several aspects, like dynamical robustness of the ground state for different perturbations, have not been explored yet. In this paper, we have performed a systematic analysis of the effect of initial velocity, quenching of the SO and Rabi coupling parameters, and allowing the collision between the components by initially perturbing the components with equal and opposite speeds. All these protocols facilitate us to obtain a variety of dynamical phases that include breathing (both in space and time), repelling, multi-fragmented solitons, etc.   

The structure of our paper is as follows. In Sec.~\ref{sec:2}, we present governing equations and numerical simulation details and outline a possible scenario to connect our numerical parameters with the experiment. We illustrate an analytical solution of the ground state of the quantum soliton in the presence of SO coupling in Sec.~\ref{sec:3}. Following this in Sec.~\ref{sec:4}, we present a detailed analysis of the ground state structure followed by its dynamics which are set up in the system by different procedures. First, we discuss the different sorts of dynamics that arise due to the initial velocity given to the soliton, followed by the dynamics due to the quenching of coupling parameters for both zero and finite speeds. Further, we highlight some of the pronounced dynamical behaviour shown by the solitons in the presence of collisions. Finally, we conclude our work in Sec.~\ref{sec:5}.

\section{Beyond Mean-field model for SO coupled BEC\lowercase{s}}
\label{sec:2}
We consider a pseudo-spin-$1/2$ Bose-Einstein condensate confined in strong transverse confinement modelled using quasi-one-dimensional spin-orbit and Rabi coupled condensates with LHY term. The corresponding coupled Gross-Pitaevskii (GP) equations in dimensionless form are given by~\cite{Tononi2019}:
\begin{subequations}
\label{eq:gpsoc:1}
\begin{align}
\mathrm{i} \partial_t \psi_{\uparrow}=\bigg[ &-\frac{1}{2}\partial_x^2-\mathrm{i} k_{L} \partial_x + g \lvert \psi_{\uparrow} \rvert^2 + g_{\uparrow\downarrow} \lvert \psi_{\downarrow} \rvert^2 \notag \\ &-\dfrac{g_{\text{LHY}}^{3/2}}{\pi}\sqrt{\lvert \psi_{\uparrow}\rvert^2 +\lvert\psi_{\downarrow}\rvert^2}\bigg] \psi_{\uparrow}+ \Omega \psi_{\downarrow}, \label{eq:gpsoc2-a} \\
\mathrm{i} \partial_t \psi_{\downarrow}=\bigg[ &-\frac{1}{2}\partial_x^2 +\mathrm{i} k_{L} \partial_x  + g_{\downarrow\uparrow} \lvert \psi_{\uparrow} \rvert^2 + g \lvert \psi_{\downarrow} \rvert^2  \notag \\
&-\dfrac{g_{\text{LHY}}^{3/2}}{\pi}\sqrt{\lvert \psi_{\uparrow}\rvert^2 +\lvert\psi_{\downarrow}\rvert^2}\bigg] \psi_{\downarrow}+ \Omega \psi_{\uparrow}, \label{eq:gpsoc2-b}
\end{align}
\end{subequations}
where $\psi_\uparrow$ and $ \psi_\downarrow$ are the wavefunctions of the spin-up and spin-down components respectively, $k_L$ is the spin-orbit coupling strength, $\Omega$ is the Rabi coupling frequency, $g$ is the intraspecies interaction, and $g_{\uparrow\downarrow}$ are the interspecies interaction strengths. We consider the interaction term due to the LHY correction as $g_{\text{LHY}}=g$. The wave functions are subjected to the following normalization condition:
\begin{align}
 \int\limits_{-\infty}^{\infty} 
\left( \lvert \psi_\uparrow \rvert^2 + \lvert \psi_\downarrow \rvert^2 \right) \, dx = 1,
\end{align}
and also that the total density $n = n_{\uparrow}+n_{\downarrow} \equiv \lvert \psi_\uparrow \rvert^2 + \lvert \psi_\downarrow \rvert^2$ is conserved.

Equations ($\ref{eq:gpsoc:1}$) are non-dimensionalized using transverse harmonic oscillator length $a_{0}=\sqrt{\hbar/(m\omega_{\perp})}$ as a characteristic length scale (where, $\omega_\perp$ is the trap frequency in the transverse direction), $\omega_{\perp}^{-1}$ as a time scale, and $\hbar\omega_\perp$ as an energy scale. The other interaction parameters are considered as $g$ = $2 N a_{\uparrow \uparrow}/ a_{0}$ and $g_{\uparrow\downarrow} = 2 N a_{\uparrow \downarrow} / a_{0}$. Here, $a_{\uparrow \uparrow}$ and $a_{\uparrow \downarrow}$ represent the scattering length corresponding to intra and inter components, respectively. The SO and the Rabi coupling parameters have been rescaled as $k_L \to k_L a_{0}$ and  $\Omega \to {\Omega}/\omega_{\perp}$, respectively, while the wave function is rescaled as $ \psi_{\uparrow, \downarrow} = \psi_{\uparrow, \downarrow} \sqrt{a_{0}}$.

To make the numerical simulation experimentally viable, we choose the parameters same as considered in the recent realization of the quantum droplet in the binary hyperfine states mixture of  $^{39}$K condensates~\cite{Cabrera2018, Cheiny2018, Ferioli2019}. Following this we consider $N\sim 10^4$ atoms confined in the harmonic trap potential with frequencies $\omega_{x} = 2\pi \times 50$\,Hz, $\omega_{\perp} = 2\pi \times 800 $\,Hz in perpendicular and axial directions, respectively. Using this the characteristic length scale can be obtained as $a_{\perp} \sim 0.6\,\mu\mbox{m}$. Generally, in the experiment two internal hyperfine states $\lvert F= 1, m_F = -1\rangle$ and $ \lvert F= 1, m_F = 0 \rangle$ are considered which can be attributed, respectively, to the pseudo-spin up $\lvert  \uparrow \rangle$, and pseudo-spin down $\lvert \downarrow \rangle$ states of our model. These two spin states have an equal number of atoms, and intra- and interspecies interaction strengths can be controlled by tuning $s$-wave scattering lengths through Feshbach resonance and by varying the magnetic field. Following the experiment we set $a_{\uparrow \uparrow} = a_{\downarrow \downarrow} = a_{\uparrow \downarrow} = 0.2686 a_{0}$ ($a_0$ is the Bohr radius) which gives the dimensionless interaction strengths as $g = g_{\text{LHY}} \approx 0.5$. Note that here we have taken all the scattering lengths to be the same to make the contribution of the mean-field term on the structure and dynamics of the soliton vanishingly small. However, the recent experiment on the quantum droplet with $^{39}K$ indicates the presence of soliton nature of the condensate is only possible when all the scattering lengths are not equal~\cite{Cheiny2018}. We expect our results will not alter much upon considering the unequal scattering lengths, as observed in the experiment. Another parameter in this system is the Rabi coupling frequency ($\Omega$), commonly used for coupling the spin states by tuning the frequency of Raman lasers ranging from $\Omega = 2\pi \times \{0.080 - 40\}\mbox{kHz}$, which implies that the dimensionless Rabi coupling frequency range as $\Omega = \{0.1, 50\}$. It is possible to vary the SO coupling strength ($k_L$) with the laser wavelength and their geometry. In the present work we have considered the dimensionless SO coupling range $k_L= \{0.1, 8\}$, where the laser wavelength varies as $\lambda_{L} = 17.8~\mu\mbox{m} - 223.2~\mbox{nm}$. Note that the LHY interaction is not equal to the MF interaction because the LHY nonlinear interaction term appears with an extra factor $\pi$. We have considered $g_{\text{LHY}} = g = 0.5$ and $1.0$, for which the quantum fluctuation term contributes approximately of the order of $11\%$ and $30\%$ respectively. Interestingly albeit of small contribution, we have obtained a large variety of dynamical features for our system without any confinement.

We employ the imaginary time propagation method with the aid of a split-step Crank-Nicolson scheme~\cite{Muruganandam2009, Young2016, Ravisankar2021} to numerically solve the coupled GP equations~(\ref{eq:gpsoc2-a}) and (\ref{eq:gpsoc2-b}). The box size $[-153.6: 153.6]$ with spatial resolution as  $dx = 0.025$ is chosen for all the simulation runs, and we use the Gaussian initial condition with anti-symmetric profiles on the components, i.e., ${\psi}_{\uparrow}(x)=-{\psi}_{\downarrow}(-x)$. The time step is fixed at $dt = 10^{-5}$. We also consider various interaction strength ($g=0.5, 1, 2$) for our studies. 

\section{Analytical soliton solution}
\label{sec:3}
\begin{figure}[!htp]
\centering\includegraphics[width=0.99\linewidth]{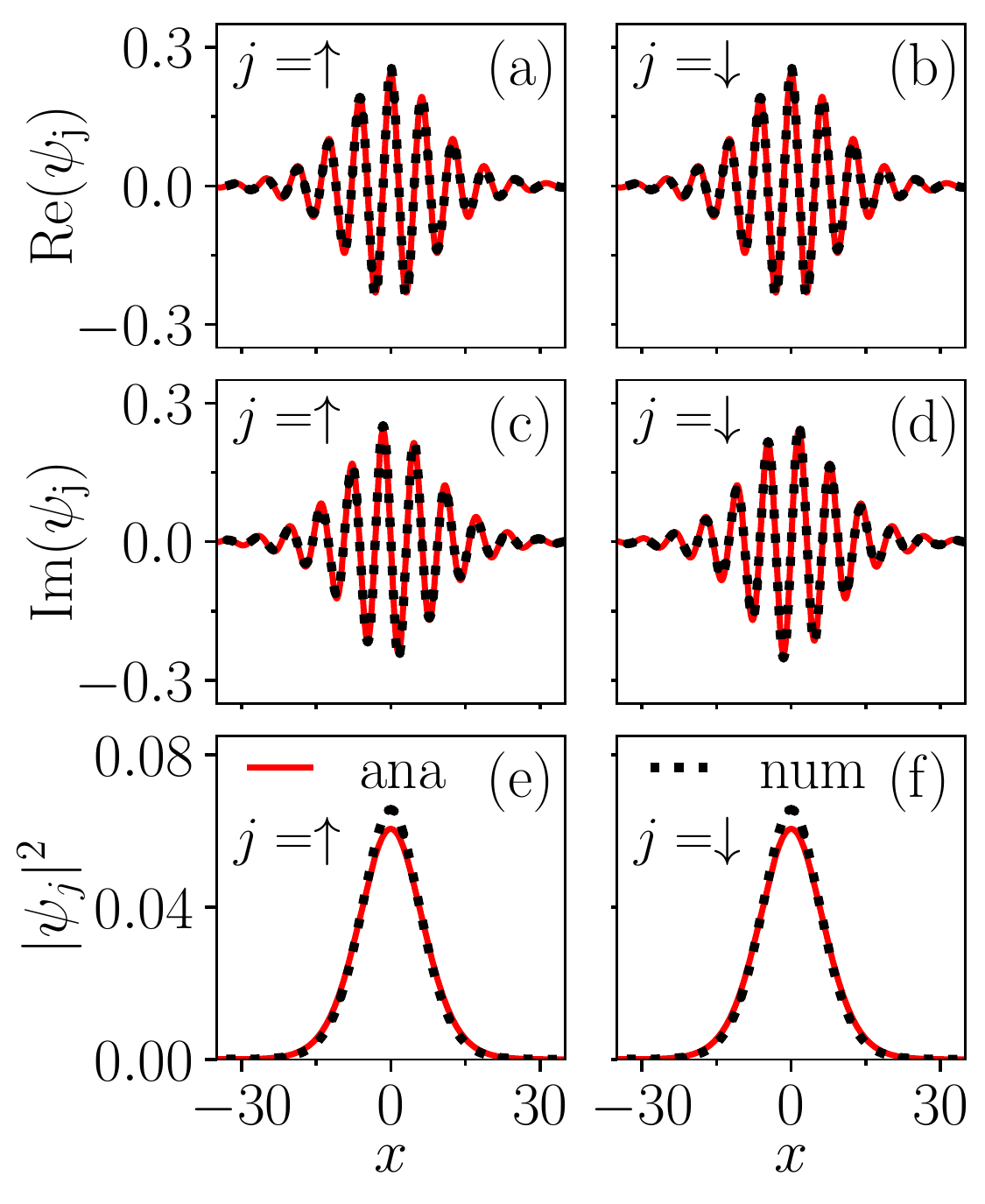}
\caption{A comparison between the numerically obtained soliton profiles (dotted-black line) and analytical (solid-red line) solution (\ref{eq:exact-ana}) for interaction strengths $g=0.5$, and coupling parameters $k_L = 1$ and $\Omega = 0$. Real parts of the spin components: (a) $\mathrm{Re}(\psi_{\uparrow})$ and (b) $\mathrm{Re}(\psi_{\downarrow})$, imaginary parts of the spin components: (c) $\mathrm{Im}(\psi_{\uparrow})$ and (d) $\mathrm{Im}(\psi_{\downarrow})$, and the densities of the (e) up $\lvert \psi_{\uparrow} \rvert^2$ and (f) down $\lvert \psi_{\downarrow} \rvert^2$ components.}
\label{fig:exact-sol}
\end{figure}

Here, we present the analytical solution of the ground state for our model. For brevity, we wish to obtain a solution for zero Rabi coupling frequency ($\Omega =0$) and finite $k_L$. For this purpose, we consider the following transformation to eliminate the SO coupling term from Eq.~(\ref{eq:gpsoc:1}),
\begin{subequations}
\label{eq:init-wave1}
\begin{align}
\psi_{\uparrow}(x,t) &= \widetilde{\psi}_{\uparrow}(x,t) \exp{\left[\frac{\mathrm{i} k_L}{2}\left(k_L t - 2 x\right)\right]} \\
\psi_{\downarrow}(x,t) &= \widetilde{\psi}_{\downarrow}(x,t) \exp{\left[\frac{\mathrm{i} k_L}{2}\left(k_L t + 2 x\right)\right]}
\end{align}
\end{subequations}
Using the single field approximation we get $\widetilde{\psi}_{\uparrow} =\widetilde{\psi}_{\downarrow}=\phi(x)\exp{\left(\mathrm{i}\mu t\right)}$. Based on the system parameters $g_{\uparrow \downarrow}=-g$, we find that the mean-field interaction term cancels each other, which yields the stationary state solution of the form  
\begin{align}
-\mu \phi = -\frac{1}{2}\partial_{x}^2\phi - \frac{\sqrt{2}}{\pi}g^{3/2} \phi^2
\end{align}
which gives 
\begin{align}\label{eq:ode}
\phi'^2 = \alpha \phi^2 -  \beta \phi^3
\end{align}
where prime ($'$) represents the spatial derivative with respect to $x$ and $\alpha = 2\mu, \beta = 2\sqrt{2}g^{3/2}/3\pi$. Now the solution of the Eq.~(\ref{eq:ode}) becomes 
\begin{align}\label{eq:sol}
\phi(x) = \phi(0)~\text{sech}^2(\sqrt{\mu/2}~x)
\end{align}
where, $\phi(0) = \alpha / \beta$. Here, the stationary state solution  depends on the chemical potential. Using the normalization condition $\int \lvert \psi \rvert^2 dx = 1$ we  obtain $2\lvert\mu\rvert = (g^2/ 3^{2/3} \pi^{4/3})$. Thus the final solution of the Eq.~(\ref{eq:gpsoc:1}) have the form as,
\begin{align}
\psi_{\uparrow} =\left(\frac{\alpha}{\beta}\right) \text{sech}^2\left(\sqrt{\frac{\mu}{2}}~x\right)\exp{\left[\frac{\mathrm{i}k_L}{2}(k_L t - 2 x) + \mathrm{i}\mu t\right]} \notag \\
\psi_{\downarrow} =\left(\frac{\alpha}{\beta}\right) \text{sech}^2\left(\sqrt{\frac{\mu}{2}}~x\right)\exp{\left[\frac{\mathrm{i}k_L}{2}(k_L t + 2 x)+\mathrm{i}\mu t\right]} \label{eq:exact-ana}
\end{align}
In Fig.~\ref{fig:exact-sol}, we plot the analytical (red line) solution [Eqs.~(\ref{eq:exact-ana})] as well numerical simulation (black dots) results of the stationary ground state of the Eq.~(\ref{eq:gpsoc:1}) for $g_{LHY}=g=0.5$, with $\Omega=0$ and  $k_L=1$. 
Both analytical and numerical simulation results match reasonably well for the real and imaginary parts and the total density of components. 

In brief, we found the approximate soliton solution for the system of coupled Eqs.~(\ref{eq:gpsoc:1}) by eliminating the SO coupling term using the transformation with zero Rabi coupling frequency. In the process, we have used single field approximation, which provides $\text{sech}^2(x)$ soliton solution due to quadratic nonlinearity~\cite{Tononi2019}. 

\section{Numerical results}
\label{sec:4}
In this section, first, we present our numerical results for different ground state solitons obtained using the imaginary-time propagation scheme. Our main emphasis is to ascertain the role of mean-field and beyond mean-field LHY term on the overall shape and structure of the self-bound quantum soliton state. Following this, we focus on the soliton dynamics by employing the real-time propagation of the governing dynamical GP equations. Finally, we present a detailed analysis of the dynamics of quantum solitons by giving initial velocity, quenching the coupling parameters and allowing the collision between the components by initially imparting them with an equal speed in the opposite direction.

\subsection{Stationary ground states of quantum solitons}
\label{sec:4a}

To better understand the role of the beyond mean-field term ($g_{\text{LHY}}$) along with the SO and Rabi coupling frequency on the shape and size of the ground state, we now proceed to get the different ground state of the solitons by assuming $g=-g_{\uparrow \downarrow} = g_{\text{LHY}}$. Here, we use MF and LHY ground states to represent ground states in the absence and presence of LHY correction, respectively. In general, depending upon the coupling parameters range, the ground states are either bright soliton (BS) or stripe soliton (SS) in nature~\cite{Tononi2019, Ravisankar2020}.

\begin{figure}[!ht]
\centering\includegraphics[width=0.99\linewidth]{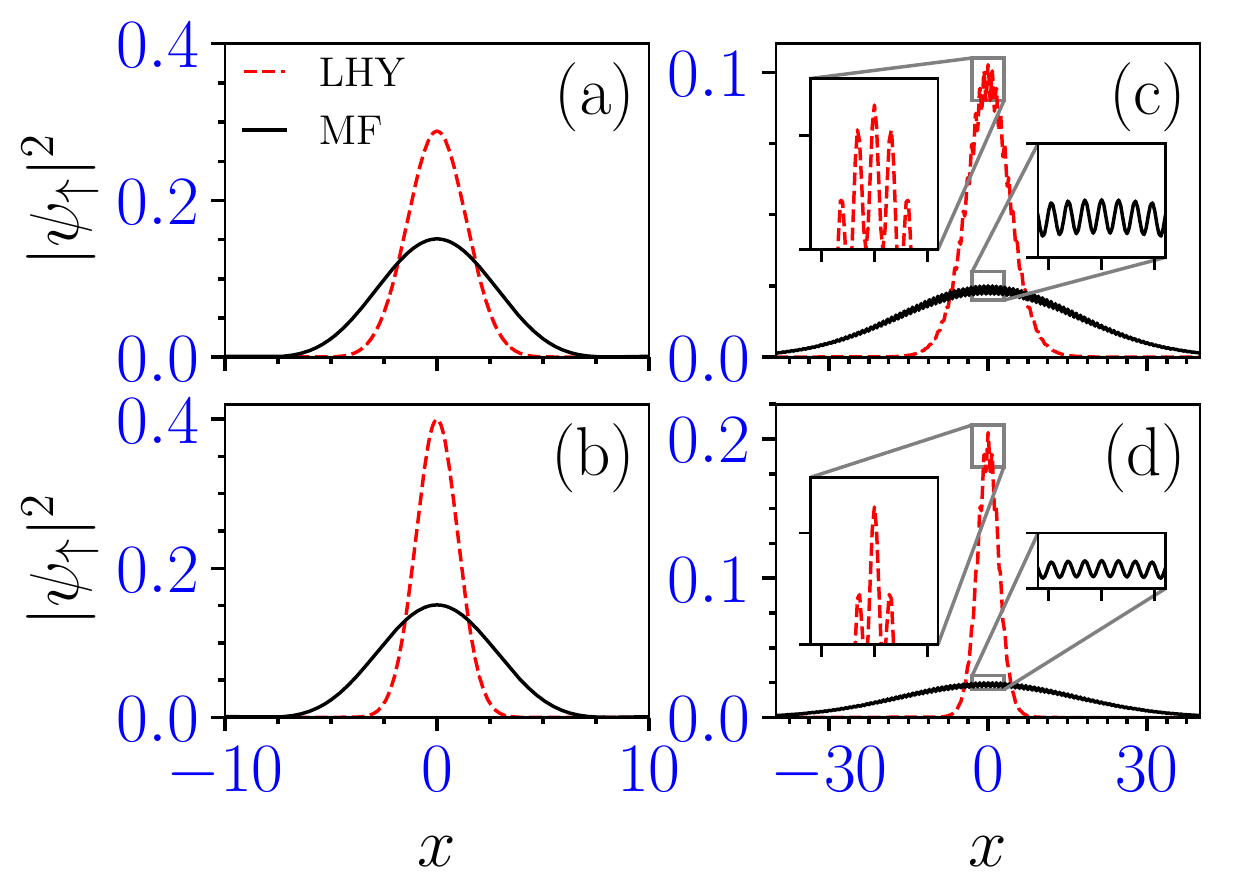}
\caption{
Ground state density profiles of the spin-up component in the presence and absence of LHY term. Bright soliton for $\Omega=1$ and $k_L=1$ with (a) $g=0.5$ and (b) $g=1.0$, and stripe soliton for $\Omega=1$ and $k_L=4$ with (a) $g=0.5$ and (d) $g=1.0$. The spin density gets more confined when the LHY term (dashed-red line) is present. The insets in (c) and (d) show the magnified views of the stripe patterns in the presence and absence of the LHY term.
}
\label{fig:denPW}
\end{figure}

We begin the analysis of the structure of BS in absence of $g_{\text{LHY}}$ for the fixed parameters $k_L= \Omega = 1$ and $g= - g_{\uparrow \downarrow} = 0.5$. In Fig.~\ref{fig:denPW}(a), the solid black line represents the BS state. In the presence of $g_{\text{LHY}}$ magnitude of the BS gets doubled, and it tends to become more localized even though the width of the soliton appears to get reduced, as represented by the dashed-red line in Fig.~\ref{fig:denPW}(a). 
\begin{figure*}[!ht]
\centering\includegraphics[width=0.85\linewidth]{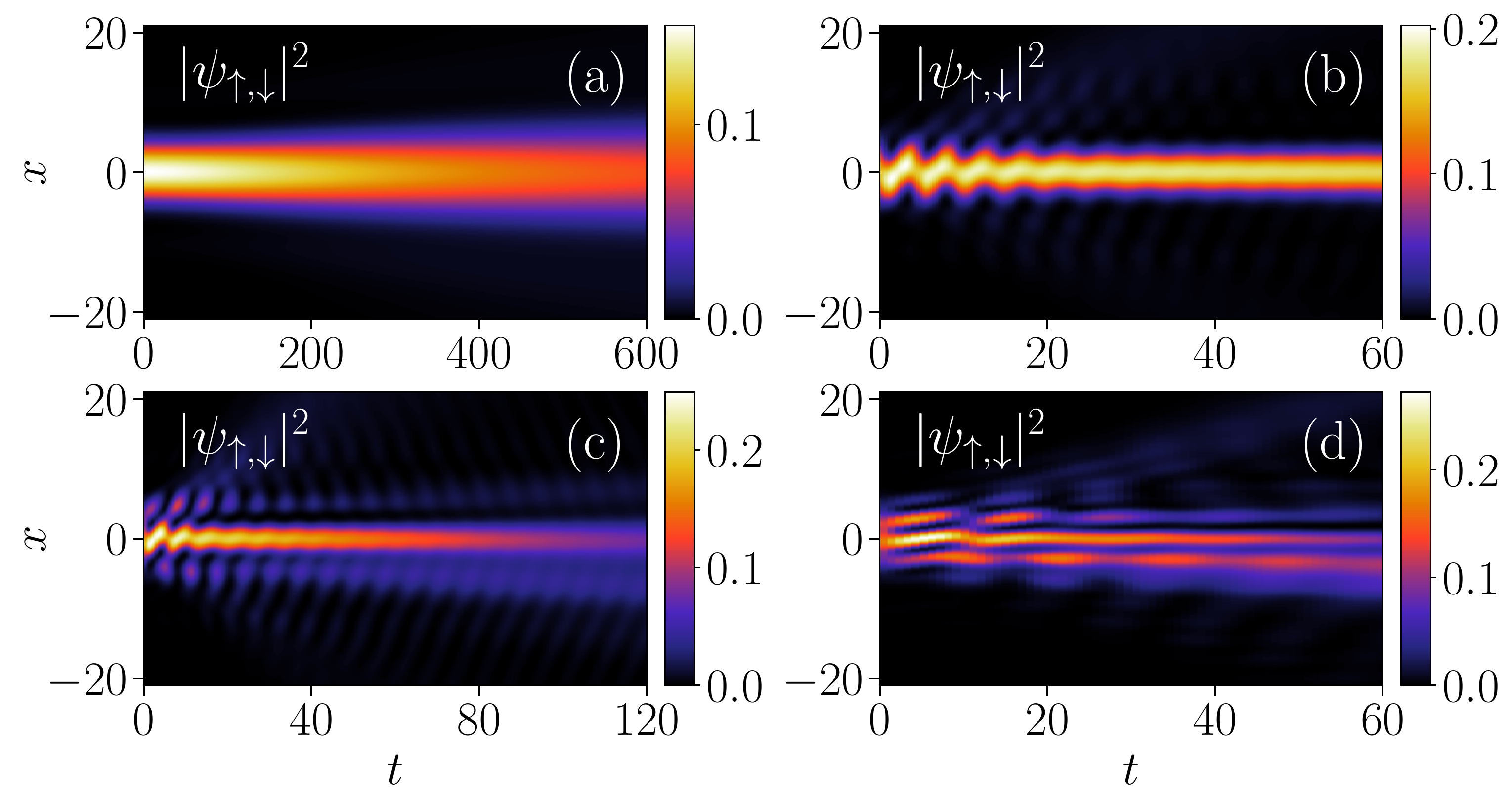}
\caption{Temporal evolution of mean-field (MF) bright soliton (BS) at different initial velocities: (a) $v=0$, (b) $v=0.2$, (c) $v=0.5$, and (d) $v=1$ for $g=-g_{\uparrow\downarrow}=0.5$ and $\Omega=k_L=1$. Upon increasing the velocity from $v=0$ to $v=0.2$, a transition from the expanding soliton to an oscillating soliton takes place. For large velocity multi-soliton (at $v=0.5$) and bifurcated  solitons (at $v=1$) are observed.}
\label{fig:dyn1}
\end{figure*}
\begin{figure}[!htb]
\centering\includegraphics[width=0.95\linewidth]{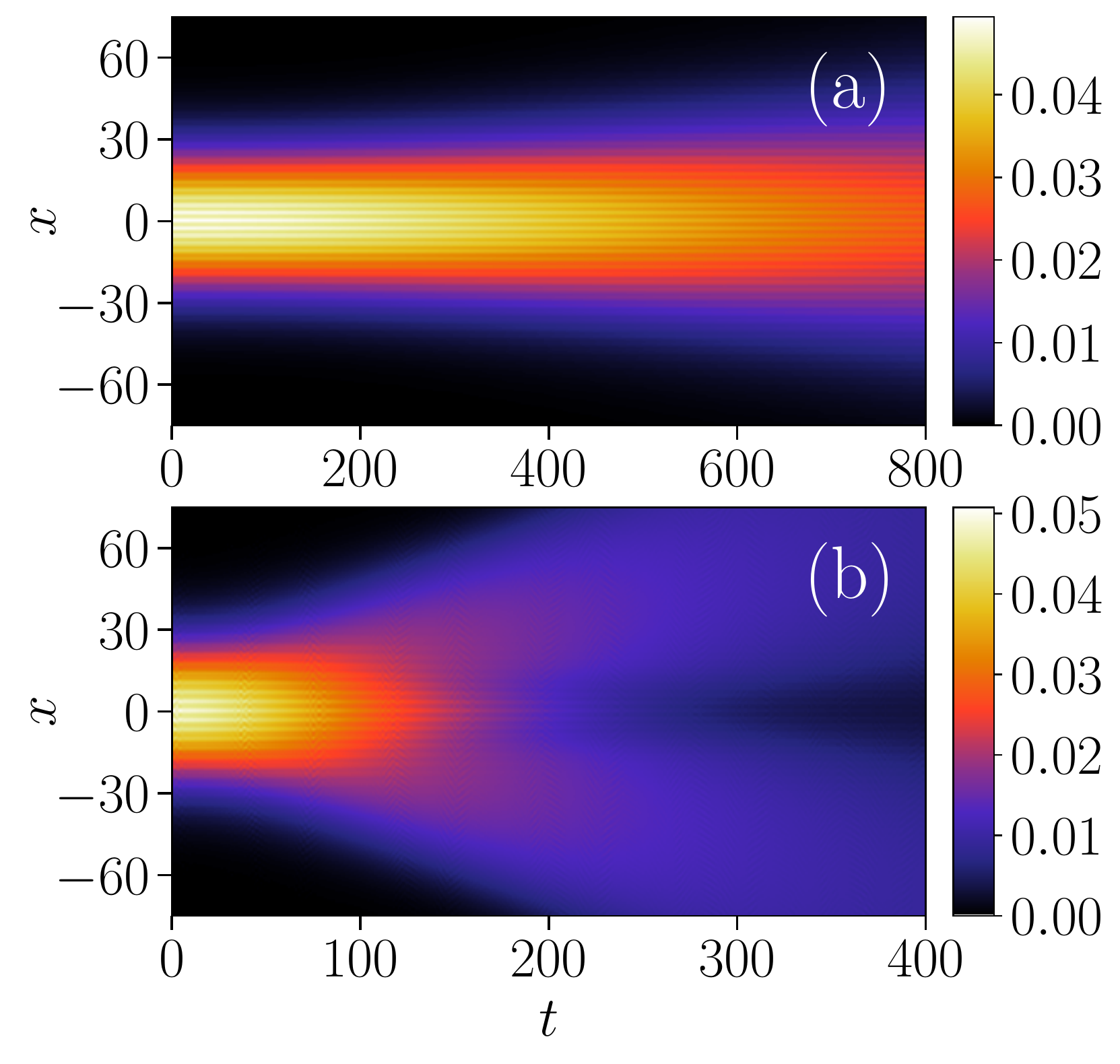}
\caption{Temporal evolution of MF stripe soliton for different velocities: (a) $v=0$ and (b) $v=0.4$ with $g=-g_{\uparrow\downarrow}=0.5$, $\Omega=1$, and $k_L=4$. For $v=0$ shape of the soliton remains unchanged until $t\sim 200$. An expansion occurs beyond this time. For finite velocity ($v=0.4$), the stripe soliton shows propagation along the initial velocity direction.}
\label{fig:dyn2}
\end{figure}
Further upon increasing the interaction strengths to $g = - g_{\uparrow \downarrow} = 1$, we find a negligible change in the amplitude and width of the mean-field Bright soliton (MF-BS) [see Fig.~\ref{fig:denPW}(b) (solid-black line)]. However, in the presence of LHY interaction ($g_{\text{LHY}} = 1$) a well localized quantum bright soliton (QBS) width gets reduced by half, while, its amplitude is increased by a factor $\approx 2.66$ [see Fig.~\ref{fig:denPW}(b) (dashed-red line)].
\begin{figure*}[!htp]
\centering\includegraphics[width=0.99\linewidth]{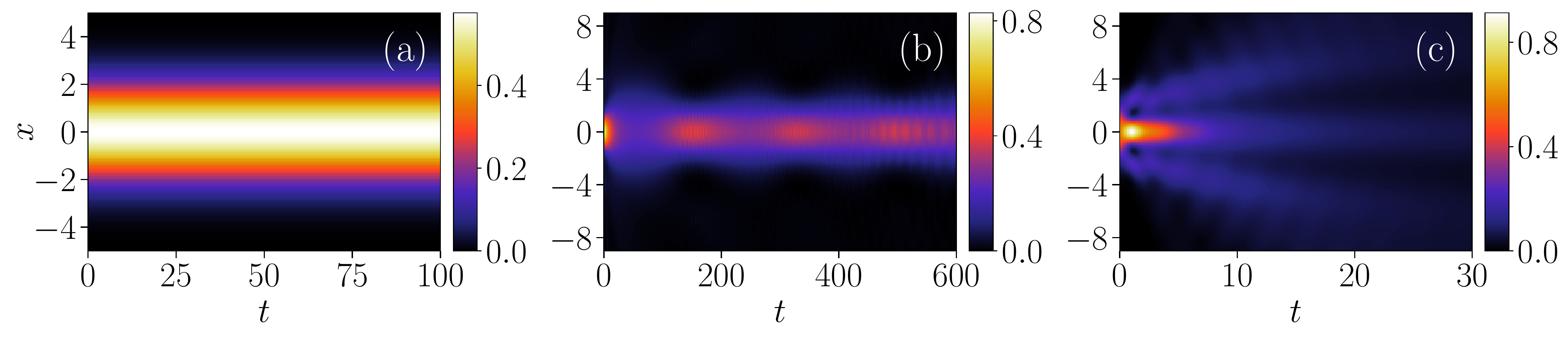}
\caption{Plots of the total density, $\sum_{j=\uparrow,\downarrow} \lvert \psi_j \rvert^2$, showing dynamics of quantum bright soliton (QBS) with LHY correction at different velocities: (a) soliton propagates with no shape change for $v=0$, (b) breathers for finite velocity $v=0.5$, and (c) oscillation and bifurcation for $v=1.0$. Increase in the velocity leads transitions from bright soliton to breather which further gets transformed into multi-solitons.
}
\label{fig:dyn3}
\end{figure*}
\begin{figure*}[!ht]
\centering\includegraphics[width=0.99\linewidth]{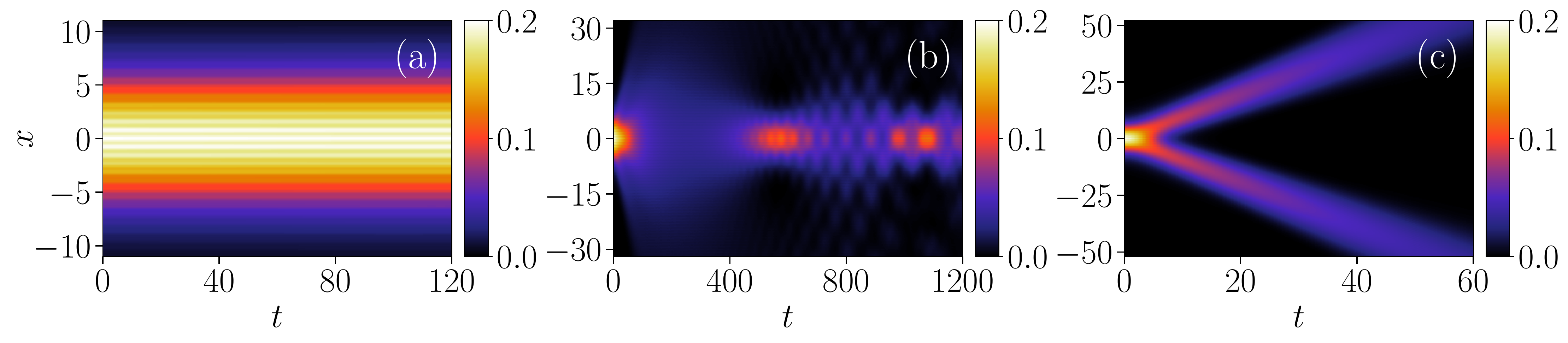}
\caption{Plots of the total density, $\sum_{j=\uparrow,\downarrow} \lvert \psi_j \rvert^2$, showing the dynamics of quantum stripe soliton (QSS) with LHY correction at different velocities: (a) $v=0$, (b) $v=0.2$, and (c) $v=1.0$  for $g=-g_{\uparrow\downarrow}=0.5$ and  $\Omega=1, k_L=4$. For $v=0$ stable stripe soliton is observed. However, for finite velocity breathers ($v=0.2$) and bifurcated ($v=1.0$) solitons are observed. } 
\label{fig:dynStrip}
\end{figure*}

Next, we consider the shape and structure of SS both in presence and absence of LHY correction term which was obtained for $\Omega = 1$ and  $k_L = 4$ at $g=0.5, 1$. A phase transition from BS to SS states occurs as we change the SO coupling strength from $k_L=1$ to $k_L=4$. The fringes that appear in the density profiles characterize the stripe wave pattern. One may note that several local maxima appear in the density profile for both the MF and quantum solitons. In the presence of the LHY term with interaction strength $g=0.5$, the number of stripes, as well as the width, gets reduced, and amplitude gets doubled compared to those without LHY, indicating the localization of the soliton [see Fig.~\ref{fig:denPW}(c)]. On increasing the interaction strength to $g=1$, we find that the mean-field stripe soliton (MF-SS) exhibits similar nature as shown with the black line in Fig.~\ref{fig:denPW}(d)]. However, the LHY-SS or quantum SS (QSS) phase width gets reduced by half, and an increase in the amplitude by a factor $\approx 4$, accompanied by a loss in the number of stripes and an increment in the localization [shown with the red line in Fig.~\ref{fig:denPW}(d)] is observed. Overall, we noticed that the condensate shape hardly changes upon increasing the MF interaction strengths. However, we note a significant change in the shape and amplitude when we consider the LHY correction ($g_{\text{LHY}}\neq0$). With the LHY correction on increasing the nonlinear interaction ($g$), the size of the soliton gets reduced while the amplitude increases. 

\subsection{Dynamics of different phases of quantum soliton}
\label{sec:4b}
In this section, we present the dynamics of the ground state of the MF and quantum soliton by solving the governing equation [cf. Eqs.~(\ref{eq:gpsoc:1})] with the help of real-time propagation. The main aim here is to investigate the dynamics of the MF and quantum solitons by giving some initial velocity to the condensate by making a uniform change in the phase of the soliton wavefunction~\cite{Denschlag2000, Tononi2019}. Further, we demonstrate breathing solitons, fragmented soliton, free expansion of soliton, etc., by manipulating the magnitude of the initial velocity.
\begin{figure*}[!ht]
\centering\includegraphics[width=0.99\textwidth]{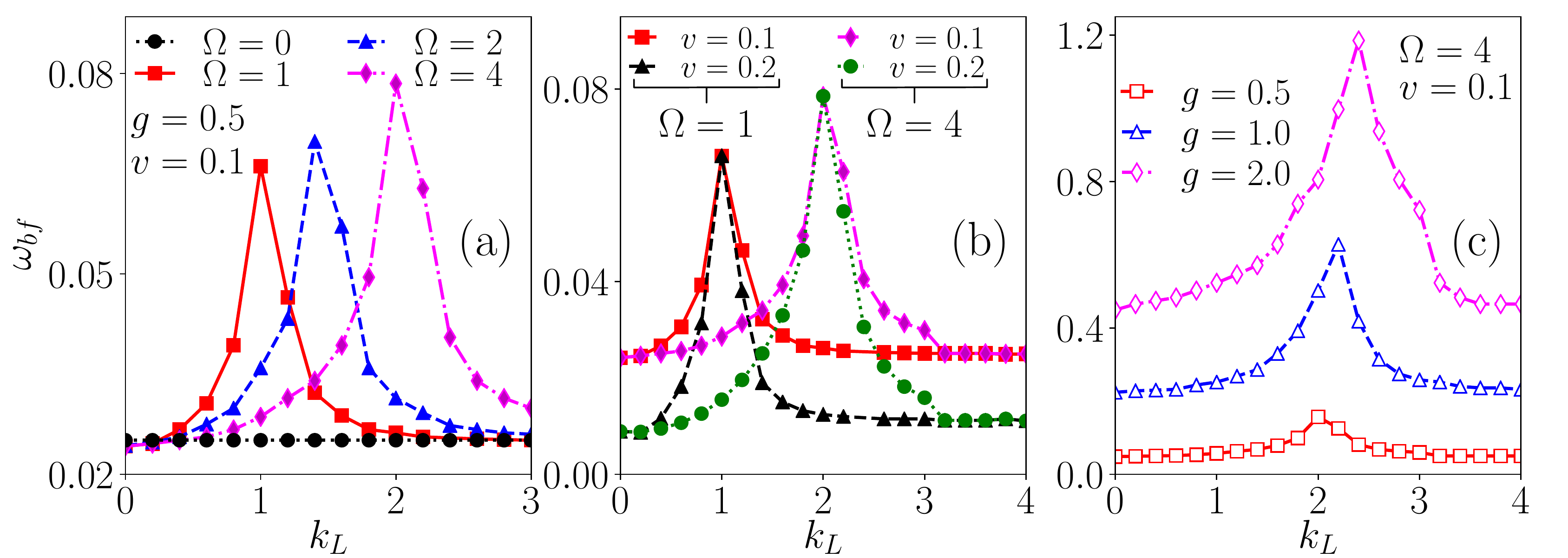}
\caption{(a) Variation of breathing frequency ($\omega_{bf}$) with the SO coupling strength ($k_L$) for different $\Omega$ with fixed velocity $v=0.1$. (b) The Variation of $\omega_{bf}$ with $k_L$ for different Rabi coupling ($\Omega=1,4$)  and velocities $v= 0.1,0.2$ with $g=0.5$. $\omega^c_{bf}$ at the critical $k_L$ where PW-SS phase transition happens and remains independent of the velocity. (c) Variation of $\omega_{bf}$ with $k_L$ for different interaction strengths $g=0.5, 1.0, 2.0$ at  $\Omega = 4$ and $v = 0.1$. To bring the graph to the same scale, we multiply the data for $g=0.5$ and $g=1.0$ by a factor of two.}
\label{fig:breath1}
\end{figure*}
Fig.~\ref{fig:dyn1} depicts the temporal evolution of MF-BS for different velocities, $v=0,0.2, 0.5$ and $1$ with $g=-g_{\uparrow\downarrow}=0.5$ and $\Omega = k_L=1$. For $v=0$, the soliton propagates without distortion until $t \sim 150$, and after this time a slight expansion is observed [see Fig.~\ref{fig:dyn1}(a)]. This feature is quite evident as the mean-field contribution vanishes, and the system does not have any trap. In this case, the soliton maintains its shape over a definite extent of time, beyond which the interactions could not stabilise the shape, resulting in the expansion indicating the appearance of dynamical instability~\cite{Ravisankar2020A}. For a small but finite initial velocity (for example, $v=0.2$), we notice setting up a time-dependent oscillation in the soliton that manifests as an undulating motion of the density in time and space. Beyond $t \sim 30$, the oscillation vanishes and stable soliton is observed [cf. Fig.~\ref{fig:dyn1}(b)]. This oscillation can be associated with the instability that soliton displays upon making a sudden change in the phase by a factor $\exp{(\pm\mathrm{i}vx)}$, where $v$ is the resultant acquired velocity. Similar to the case $v=0$, we observe an expansion in the soliton upon waiting for a longer period.  For $v=0.5$, soliton appears to display breathers-like oscillation in which it exhibits expansion followed by compression at a periodic time interval [see Fig.~\ref{fig:dyn1}(c)]. The spin component exhibits interesting dynamical behaviour. We noticed the absence of the breather on one side while it is present on the other side. However, the middle lobe exhibits oscillation for a shorter time and expansion in a long duration indicates the presence of an oscillation death-like behaviour in the soliton~\cite{Ravisankar2020A}. In addition, we also observe the presence of multi-soliton-like features. At $v=1$, fragmented solitons with breather-like modes get generated and propagate in the respective directions for the spin components, accompanied by the appearance of bifurcation-like solitons [cf. Fig.~\ref{fig:dyn1}(d)].

We observe similar nature of dynamical expansion in the SS as those observed for BS in the absence of the LHY term. In Fig.~\ref{fig:dyn2} we show the temporal evolution of SS in absence of LHY term for $\Omega=1$ and $k_L=4$ with $g=0.5$ for (a) $v=0$  and  (b)$v=0.4$. For $v=0$, soliton does not have a stable structure and exhibits expansion for a long time. However, at higher velocity  ($v=0.4$), bifurcation of the stripe soliton into two propagating solitons occurs, which exhibits expansion with time.

After discussing the dynamics of the soliton in the presence of MF term, we now analyse the dynamical evolution when we consider the LHY term for both QBS and QSS. In Fig.~\ref{fig:dyn3}, we illustrate the dynamics of QBS with the LHY correction term for different initial velocities $v=0,0.5$ and $v=1.0$ at $g=-g_{ab}=0.5$ and $\Omega=k_L=1$. For zero velocity soliton exhibits stable nature with no change in its shape and size with time as depicted in Fig.~\ref{fig:dyn3}(a). The MF dynamical instability gets under control due to the presence of $g_{\text{LHY}}$ interaction, stabilsing the expanding condensate without any trap. On increasing the velocity to $v=0.5$, the soliton displays breathers-like expansion and compression behaviour at periodic intervals, as shown in Fig.~\ref{fig:dyn3}(b). The frequency of these breathing modes depends on the coupling parameters, which we shall discuss in detail in the later part of the paper. Note that Tononi et al. reported similar breathers like soliton~\cite{Tononi2019}, which attained by a slight perturbation in the width of the ground state. As we increase the initial velocity to $v=1$, the soliton develops periodic breather oscillations and exhibits bifurcation-like features in which the centre of the soliton moves in the $\pm x$ directions. Note that the up-spin component moves in the $-x$ direction while the other component moves in the $+x$ direction. This feature is noticeable for a finite time $t\lesssim 10$. Beyond this time, the breathers expand and span the entire space while the centred soliton exhibits expansion. Apart from this, we also notice that it exhibits bright-dark soliton for a shorter time, which finally gets converted into multi-soliton at a longer period~\cite{Astrakharchik2018} as displayed in Fig.~\ref{fig:dyn3}(c). 

Further, we focus on the dynamics of QSS with LHY correction at different velocities $v=0,0.2,$ and $1.0$ for the parameters $g=-g_{ab}=0.5$, $\Omega=1$, and $k_L=4$. At $v=0$, QSS remains stable and maintains its shape and size for a longer duration [see Fig.~\ref{fig:dynStrip}(a)]. However with finite velocity ($v=0.2$) QSS propagates for a while in the $\pm x$ directions, thereafter, for $t\gtrsim 50$ they start displaying decaying  (dilatation) feature until $t \lesssim 300$. Following this the dilatation slowly gets diminished and for $t\gtrsim500$ soliton gets revived [see Fig.~\ref{fig:dynStrip}(b)]. The whole revival and expansion of the soliton with time is similar to those obtained for the QBS [see Fig.~\ref{fig:dyn3}(b)]. For $v>0.5$, the QSS evolves with time in the respective directions, and after separation, they do not combine again, as shown for $v=1.0$ in Fig.~\ref{fig:dynStrip}(c).

\begin{figure}[!ht]
\centering\includegraphics[width=0.49\textwidth]{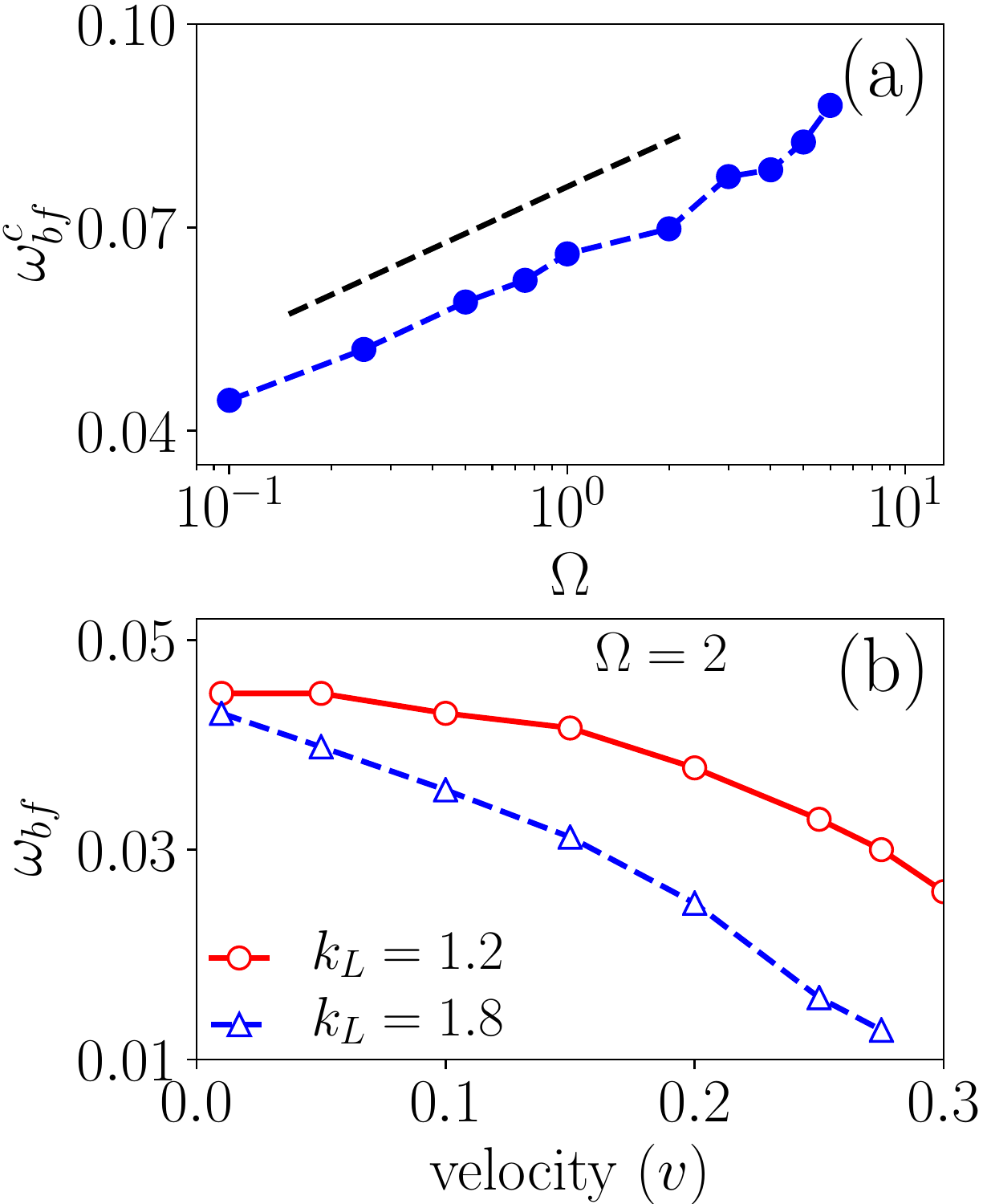}
\caption{(a) Variation of breathing frequency ($\omega^c_{_{bf}}$) at critical $k_L$ with  $\Omega$ for fixed velocity $v=0.1$. The other parameters are the same as those in Fig.~\ref{fig:breath1}. The critical breathing frequency increases and exhibits power law behaviour with $\Omega$ as $\omega^c_{_{bf}}\sim \Omega^{0.16}$. The dashed line is drawn as a guide to the eye to show the power law nature of the $\omega^c_{_{bf}}$ with $\Omega$. (b) Variation of breathing frequency $\omega_{bf}$ with initial velocity $v$ for different $k_L$ keeping Rabi frequency fixed to $\Omega=2$. The $\omega_{bf}$ is small for higher $k_L$ for all the velocity.}
\label{fig:breath1a}
\end{figure}

Next, we analyze the dependence of the breathing frequency for the quantum soliton as observed in Figs.~\ref{fig:dyn3}(b) and ~\ref{fig:dynStrip}(b) on the SO and Rabi coupling parameters. We compute the breathing frequency ($\omega_{bf}$) using the time series of the density oscillation of the soliton. In Fig.~\ref{fig:breath1}(a) we show dependence of $\omega_{bf}$ on the SO couplings ($k_L$) for various $\Omega$  when the soliton was assigned the initial velocity as $v=0.1$. Following Ref.~\cite{Tononi2019}, here we consider the initial wavefunction as Gaussian, which is different than that of the exact solution of the ground state. 
The breathing frequency is around $0.025$ for $k_L=0$ for all values of $\Omega$. At $\Omega=0$ the $\omega_{bf}$ remains unchanged upon increasing $k_L$. For finite Rabi coupling frequency ($\Omega = 1, 2, 4$), breather starts appearing at periodic intervals. For fixed $\Omega$ as $k_L^2 < \Omega$ the breathing frequency increases upon increasing in $k_L$. It attains the maximum at critical SO coupling ($k^2_L=\Omega$) where the QBS to QSS phase transition occurs. For  $k_L^2 > \Omega$, $\omega_{bf}$ decreases slowly in the regime of quantum stripe soliton. At higher $k_L$,  $\omega_{bf}$ have the same value as those at $k_L\sim0$ for all $\Omega$. The critical breathing frequency ($\omega^c_{bf}$) increases upon an increase in the Rabi frequency. Fig.~\ref{fig:breath1}(b) illustrates the variation of $\omega_{bf}$ with $k_L$ for different velocities, $v= 0.1$ and $0.2$, at  $g=0.5$. We have considered two sets of Rabi couplings $\Omega = 1$ and $4$ for the analysis. The overall variation in $\omega_{bf}$ with $k_L$ appears to be the same for both the velocities ($v= 0.1$ and $0.2$), except that at lower velocity ($v=0.1$) it has higher value compared to those at large velocity ($v=0.2$). Interestingly, $\omega^c_{bf}$ remains the same for all the velocities ($v=0.1$ and $0.2$) as shown in Fig.~\ref{fig:breath1} (b), indicating that the breathing frequency at the critical point remains unchanged upon increasing the velocity. %
To better understand the dependence of the $\omega_{bf}$ on the interaction strengths in Fig.~\ref{fig:breath1}(c) we show its variation with $k_L$ for different $g$ ($g=0.5, 1.0, 2.0$)  at $\Omega = 4$ and $v = 0.1$.  We observe that the $\omega_{bf}$ for a given $k_L$ shows increasing behaviour on increasing the interaction strength. The critical SO coupling ($k^c_L$)  and corresponding critical breathing frequency ($\omega^c_{bf}$) also increase upon the increase in $g$. %

In Fig.~\ref{fig:breath1a}(a), we plot the variation of $\omega^c_{bf}$ with $\Omega$ for $v=0.1$. We find that $\omega^c_{bf}$ shows power law dependence on $\Omega$ with an exponent, $0.16$. In Fig.~\ref{fig:breath1a}(b), we plot the variation of $\omega_{bf}$ with initial velocity for fixed Rabi coupling ($\Omega = 2$) and for two sets of SO couplings $k_L=1.2$ (solid red line) and $k_L=1.8$ (dashed blue line). Here, we notice that $\omega_{bf}$ decreases upon increasing the initial velocity for both the $k_L$. The fall in the breathing frequency with the velocity becomes sharper upon the increase in $k_L$. However, for a given initial velocity, the $\omega_{bf}$ is small for $k_L=1.8$, which happens to be in QSS than those for $k_L=1.2$ at which we have QBS ground state. We note that the chemical potential in the QBS regime does not change with $k_L$ until $k_L<k^c_L$, which indicates that the soliton is energetically stable in the BS regime. On the other hand, when we increase the Rabi frequency, $\Omega$ to $2$, the critical $k_L$ for the transition from the bright to stripe soliton also increases. Overall we find that the ground state energy of the quantum soliton gets lower upon increasing $\Omega$ for a given $k_L$, suggesting a better way of obtaining an energetically stable soliton.


Another notable feature is the transition of the breathing soliton phase into the moving soliton upon the increase in the velocity as depicted in Figs.~\ref{fig:dyn3} and \ref{fig:dynStrip}. We can understand the instability that leads to the transition from breathers to the moving soliton by looking at the SO, Rabi and the kinetic part of the chemical potential. While the kinetic term varies as $v^2$, the SO and Rabi part will have linear and square dependence on the velocity respectively. Due to this, one witnesses competition between the kinetic energy, which is positive and the SO and Rabi parts will have negative values upon changing the velocity. Overall, we expect the domination of the breathing soliton region by the SO and Rabi coupling contribution, while the moving soliton region kinetic energy dominates. The above also indicates that the chemical potential attains minimum at the critical velocity, which is consistent with our numerical observation.

\color{black}
After analyzing the dynamics of the solitons by imparting an initial velocity, in the following section, we present different dynamical features arising due to making an instantaneous quench in the coupling parameters ($\Omega$ and $k_L$). 


\subsection{Quench dynamics of quantum soliton}
\label{sec:4c}
To better understand the effect of the SO and Rabi couplings on the overall dynamics (without giving any initial velocity) of the quantum solitons in this section, we employ the instantaneous quench of the coupling parameters and analyze the resultant dynamics. With this protocol, we have obtained a variety of notable features like the generation of secondary and repulsive solitons, the dynamical phase transition from the QBS to the QSS and vice versa, etc.
\begin{figure*}[!htb]
\centering\includegraphics[width=0.99\linewidth]{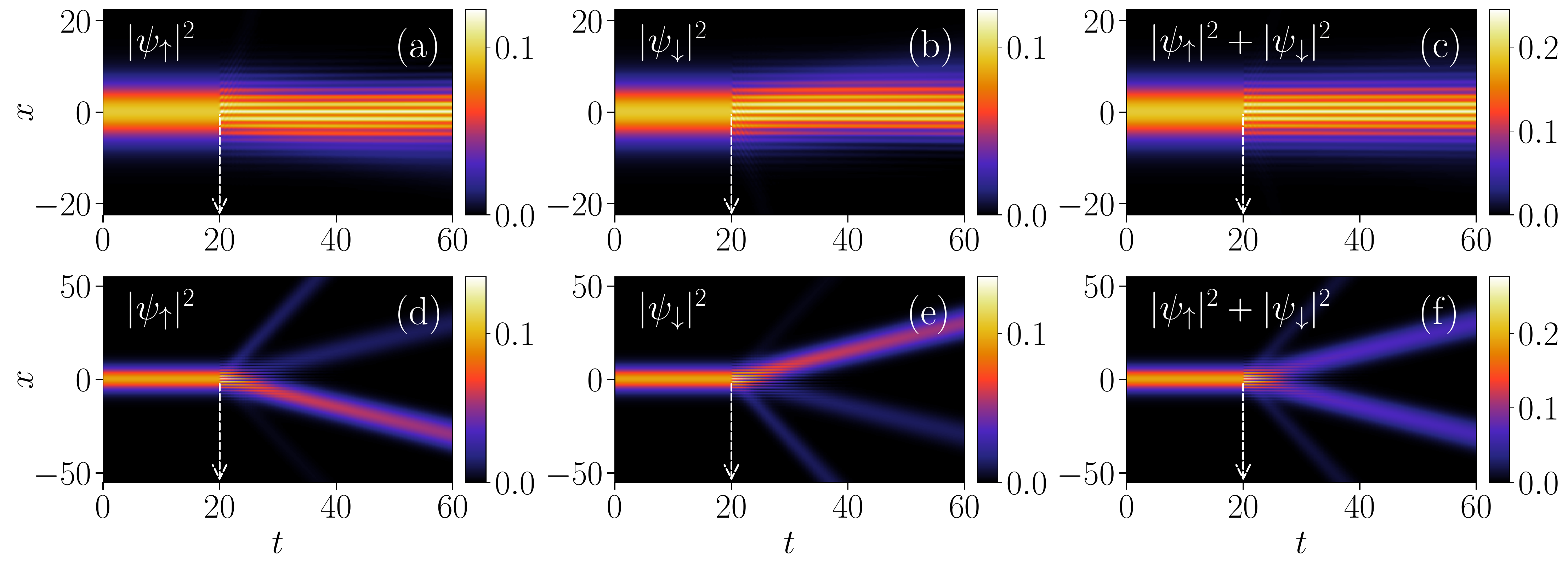}
\caption{Dynamics of the quantum soliton appearing due to different quenching protocols at $g = - g_{\uparrow\downarrow} = 0.5$ as it was initially prepared for $\Omega = 0, k_L = 2$. (a)-(c): When the Rabi frequency is quenched as  $\Omega = 0 \to 1$ at $t=20$, the initial QBS phase changes to QSS, and (d)-(f): Quenching of Rabi frequency as $\Omega = 0 \to 5$ at $t=20$ transforms the QBS phase into repulsive solitons.}
\label{fig:quench:2a}
\end{figure*}

\begin{figure*}[!ht]
\centering\includegraphics[width=0.99\linewidth]{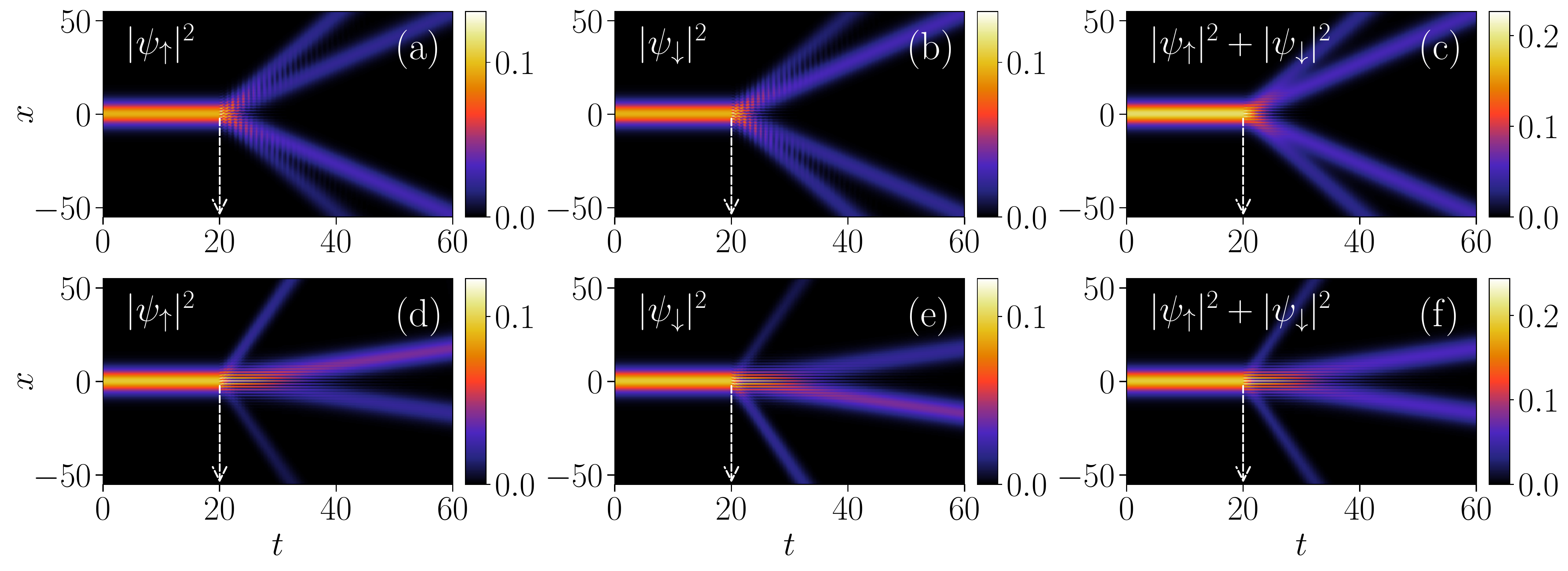}
\caption{Dynamics of the quantum soliton appearing due to different quenching protocols at $t=20$ with $g = - g_{\uparrow\downarrow} = 0.5$ as it was initially prepared for $\Omega = 0$, $k_L = 2$. (a)-(c): When both Rabi frequency and SO coupling are quenched as $\Omega = 0 \to 50, k_L = 2 \to 4$ at $t=20$, which results the transformation of non-moving QBS soliton into moving multi soliton with large repulsion angle. (d)-(f): For quenching  $\Omega = 0 \to 50, k_L = 2 \to 8$ at $t=20$, QBS undergoes bifurcation into moving multi soliton state with small repulsive angle.}
\label{fig:quench2b}
\end{figure*}

In Fig.~\ref{fig:quench:2a}, we show the dynamical evolution of the soliton which generally arises due to the quench of Rabi coupling frequency by fixing the interaction strength to $g = - g_{ab} = 0.5$. The ground state soliton was initially prepared for $\Omega = 0, k_L = 2$. As we quench the Rabi frequency from $\Omega = 0 \to 1$  at finite time ($t\sim 20$) we notice that the QBS gets transformed to the QSS [see Figs.~\ref{fig:quench:2a}(a-c)]. However, upon employing  $\Omega = 0 \to 5$, the QBS phase of the soliton gets transformed into a phase that displays expansion and certain characteristics of the repulsive soliton as shown in Fig.~\ref{fig:quench:2a}(b-d). 

\begin{figure*}[!ht]
\centering\includegraphics[width=0.99\linewidth]{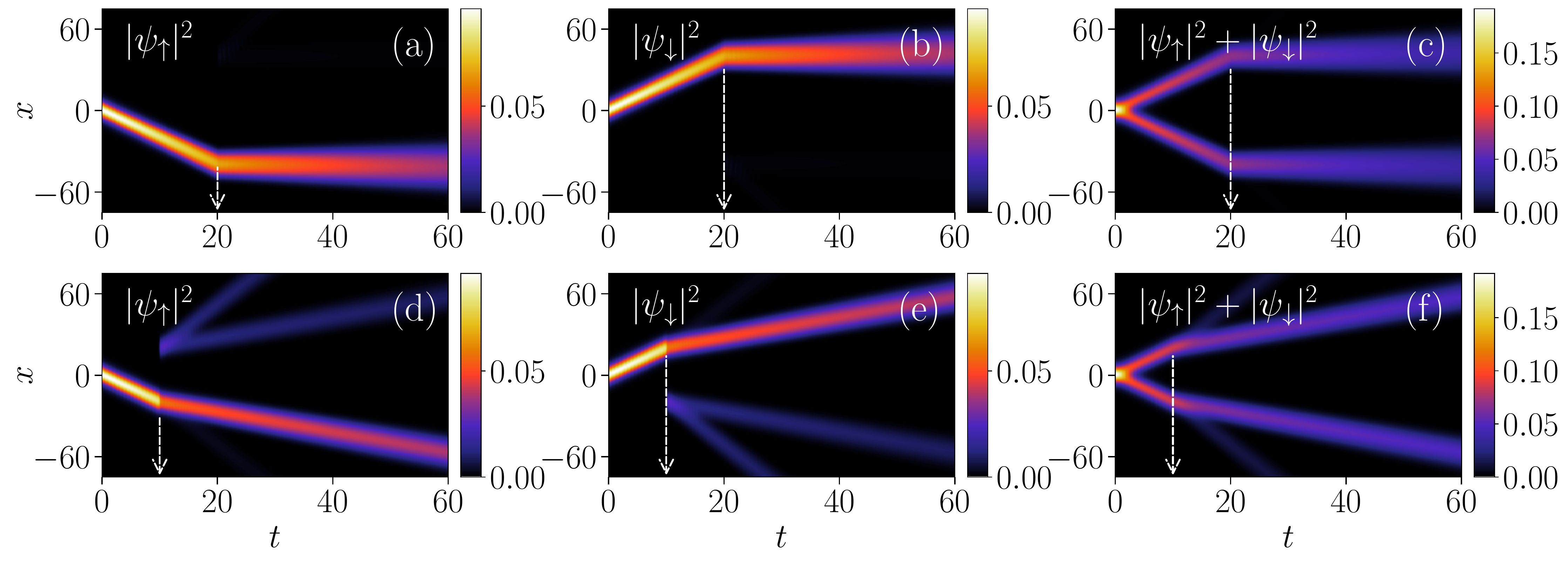}
\caption{Dynamics of the quantum soliton prepared with $g = - g_{\uparrow\downarrow} = 0.5$, $\Omega = 0$, and $k_L = 2$ appearing due to double quenching protocols. (a)-(c): first quenching is performed at $t = 0$ with change in SO coupling ($ k_L = 2 \to 0$)  and second quenching is done at $t=20$ when both coupling parameters are changed as $\Omega = 0 \to 1$ and $k_L=0\to 2$ which results as transforming the moving soliton to the stationary soliton. (d)-(f): At $t=0$ first quenching is same as those in (a)-(c), while, at $t=10$ second quenching is done as  $\Omega = 0 \to 5$ that result transforming the moving soliton into soliton with secondary waves.}
\label{fig:quench2b_case2}
\end{figure*}

\begin{figure*}[!ht]
\centering\includegraphics[width=0.99\linewidth]{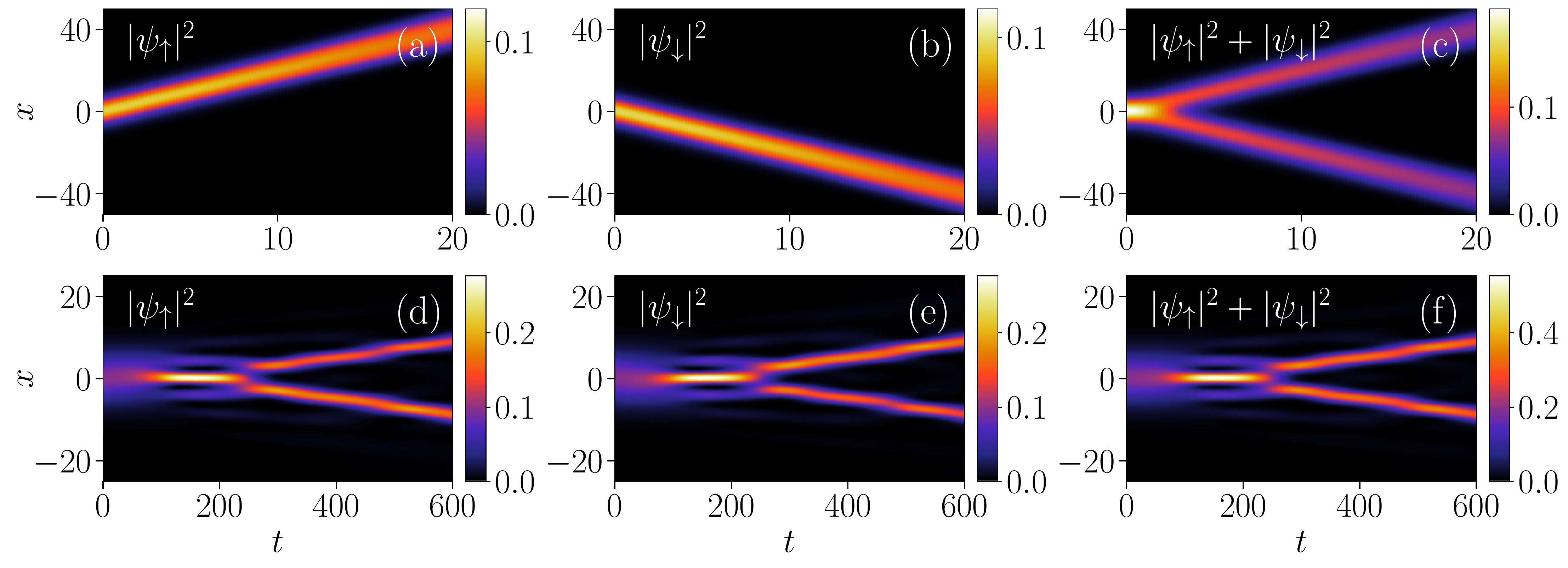}
\caption{Dynamics of the quantum PW soliton prepared with $g = - g_{\uparrow\downarrow} = 0.5$ and $\Omega =2$, $k_L = 0$ as quenching of both the coupling parameters are performed. (a)-(c): Quenching is done as $k_L = 0 \to 2$, $\Omega = 2 \to 0$ that results the transition of bright soliton into moving soliton. (d)-(f): With quenching $k_L = 0 \to 2$, $\Omega = 2 \to 4$ breather soliton gets bifurcated into filament like soliton beyond $t\sim250$.}
\label{fig:quench2b_case2a}
\end{figure*}


In Fig.~\ref{fig:quench2b}, we show the dynamical evolution of the solitons for two kinds of quenching protocols: (a-c) when $\Omega = 0 \to 50$ and  $k_L = 2 \to 4$ and (d-f) when $\Omega = 0 \to 50, k_L = 2 \to 8$. In the first case, we find the appearance of multisolitons with interference-like patterns. Following the quenching, solitons get bifurcated into four that propagate away from each other and behave like repelling solitons. In the density profile, the outermost solitons show breathing-like features, and the innermost one inherits pure, stable soliton-like characteristics, i.e., they do not exhibit any change in shape and size with time as shown in Fig.~\ref{fig:quench2b}(a-c). For the second quenching protocol ($\Omega = 0 \to 50$ and $k_L = 2 \to 8$), again, multisoliton behaviour appears, which has some differences compared to the previous case. For this case, we notice that the outermost soliton-breather gets transformed into solitons with a relatively larger repulsive angle, while the innermost solitons transform into the stripe solitons and show attraction towards each other as depicted in Fig.~\ref{fig:quench2b}(d)-(f). Thus, quenching of the coupling parameters (either one of Rabi and SO or both simultaneously) leads to secondary solitons.

\begin{figure*}[!ht]
\centering\includegraphics[width=0.99\linewidth]{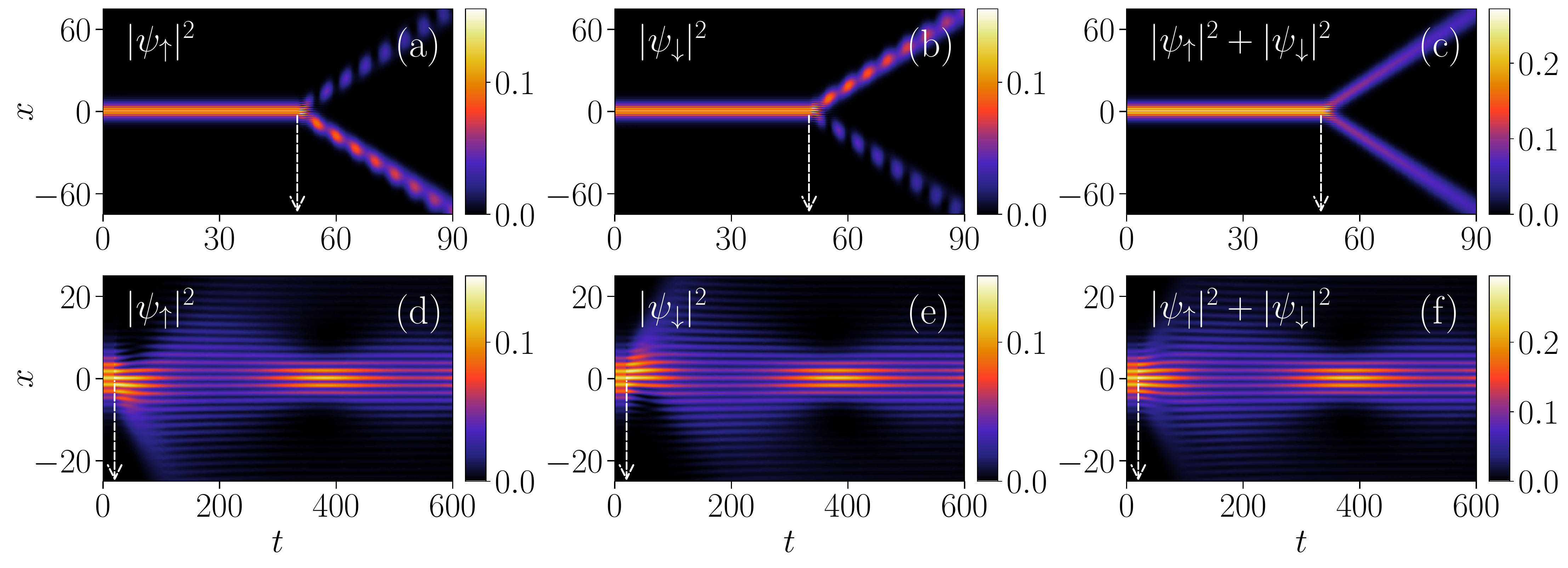}
\caption{Dynamics of the quantum stripe soliton prepared with $g = - g_{\uparrow\downarrow} = 0.5$ and $\Omega =1, k_L = 2$ and the quenching is performed at finite time. (a)-(c): Quenching is performed as $k_L = 2 \to 0.2, \Omega =1 \to 0.5$ at $t=50$ that results the transition from the stripe soliton to the space-time breather bright soliton.  (d)-(f): Quenching is performed as $\Omega =1 \to 2, k_L = 2$ at $t=20$ resulting the transition from the stripe soliton to the breathing stripe soliton.}
\label{fig:quench2b_case3}
\end{figure*}

Next we discuss the effect of double quench protocols on the dynamics of the soliton. In Fig.~\ref{fig:quench2b_case2} we show the dynamics of the quantum solitons as we implement double quenching protocols~\cite{Li2019} of the coupling parameters for the initial ground state prepared at $\Omega = 0, k_L = 2$ for $g = - g_{\uparrow\downarrow} = 0.5$. In Fig.~\ref{fig:quench2b_case2}(a)-(c), we show the evolution when first quench is performed by instantaneously switching off the SO coupling, that is, making the change $ k_L = 2 \to 0$ at $t = 0$. Subsequently soliton is allowed to evolve until $t=20$ when another quench was executed by changing Rabi frequency as $\Omega = 0 \to 1$ and switching on the SO coupling as $k_L = 0 \to 2$. The first quenching in which the SO coupling is switched off initially allows the solitons corresponding to the up and down components to display repulsive behaviour and move in the space until the second quenching is done. After the second quench  ($\Omega = 0 \to 1, k_L = 0 \to 2$) at $t=20$, we notice a switching off the soliton position from $x=0$ to some finite value which further remains invariant with time. Overall we see that the double quench protocol provides more stability to the solitons. This particular feature is one of the promising potential application for quantum information and quantum computing, which based on mixing and demixing of qubits~\cite{Mardonov2015, Ravisankar2021b}. As we perform the first quenching same as previous case while in second quench only change in Rabi frequency is executed with $\Omega = 0\to5$ at $t=10$ we find the appearance of moving BS with generation of the second harmonics [see Fig.~\ref{fig:quench2b_case2}(d)-(f)].

Next we consider $\Omega = 2, k_L = 0 $ with $v=0$ which shows only the stable BS. As we quench $k_L = 0 \to 2$ and $\Omega = 2 \to 0$ the soliton repel each other [see Fig.~\ref{fig:quench2b_case2a}(a)-(c)].  Upon quenching the Rabi coupling frequency from $\Omega = 2 \to 4$ we find that the soliton gets transformed into filaments after a finite time  ($t \gtrsim 150$), which also appears to repel each other at ($t \gtrsim 250$) as depicted in Fig.~\ref{fig:quench2b_case2a} (d)-(f). 


\begin{figure}[!ht]
\centering\includegraphics[width=0.99\linewidth]{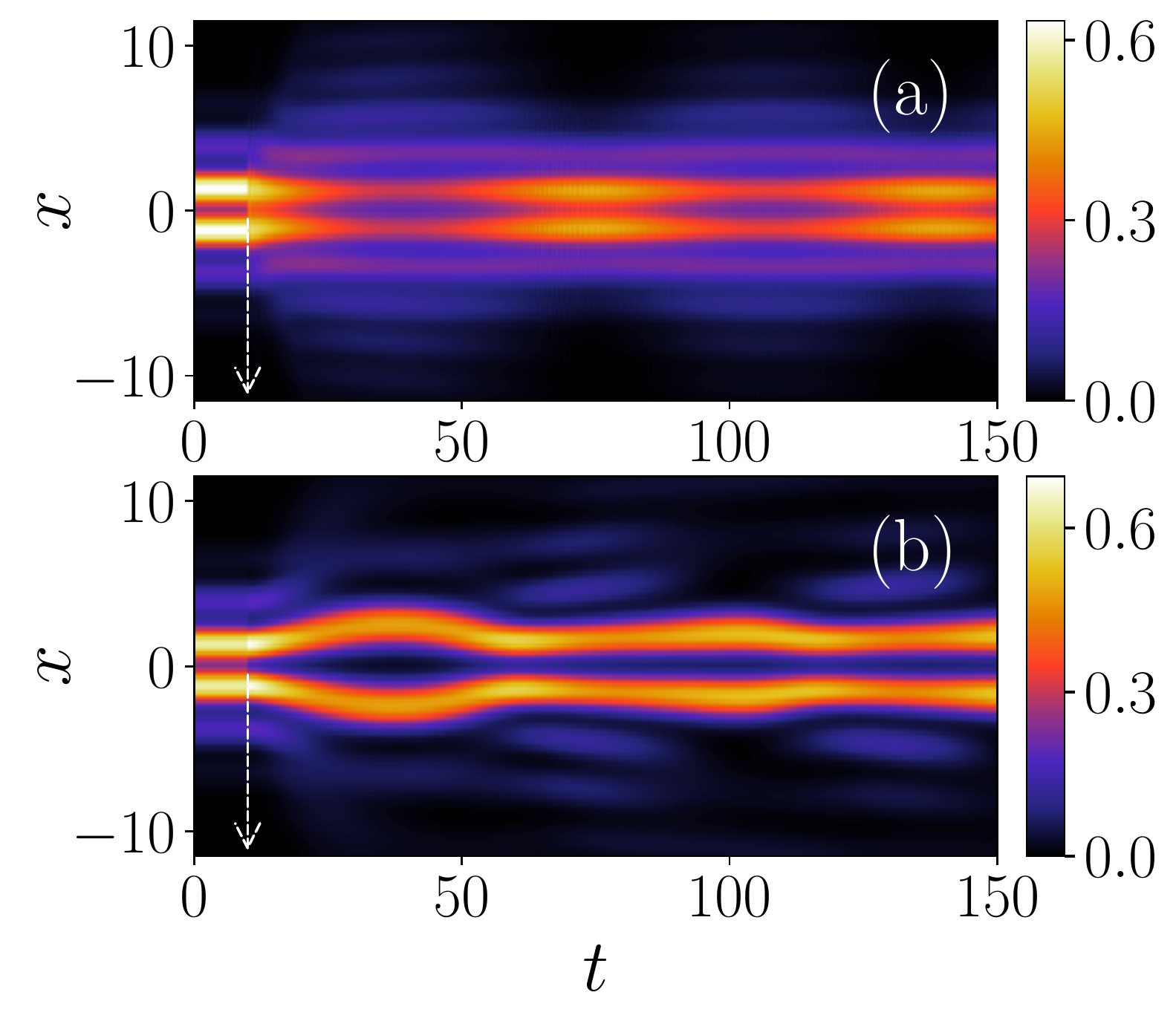}
\caption{Dynamics of quantum stripe soliton prepared with $g = - g_{\uparrow\downarrow} = 1.0$  and  $\Omega = k_L = 2$ as the quenching is performed on the Rabi frequency at $t=10$. (a) Soliton gets bifurcated into repulsive and attractive solitons as Rabi coupling is quenched as (a) $\Omega = 2 \to 1$ and (b)  $\Omega = 2 \to 3$. }
\label{fig:quench2b_case4}
\end{figure}

After discussing different dynamical behaviour arising in the quantum bright soliton due to the quench of the coupling parameter in what follows, we present the related dynamics for the quantum stripe soliton. In Fig.~\ref{fig:quench2b_case3}, we depict the dynamics that arise due to the quenching of both coupling parameters when the ground state is quantum stripe soliton nature. 


In Fig.~\ref{fig:quench2b_case3}(a)-(c) we show the dynamics of soliton when quenching of both coupling parameters are performed at $t=50$. The quenching protocol for SO coupling is implemented as $k_L = 2 \to 0.2$ while for Rabi coupling frequency it is $\Omega =1 \to 0.5$. We noticed that as a result of the quenching, stripes solitons get bifurcated into two parts wherein one shows weak breather-soliton, while the other exhibits stronger behaviour. Further decrease of Rabi coupling frequency leads to the disappearance of breathers and the generation of repulsive solitons. In Fig.~\ref{fig:quench2b_case3}(d)-(f) we depict the dynamics of soliton when quenching of Rabi coupling frequency is executed from $\Omega =1 \to 2$ at $t=20$. The quenching generates the stripe soliton breathers. Similar kinds of breathing solitons have been observed upon quenching the interaction strength so that the mean field contribution becomes quite small in the binary mixture of the condensates~\cite{Mistakidis2021}. 

\begin{figure*}[!ht]
\centering\includegraphics[width=0.99\linewidth]{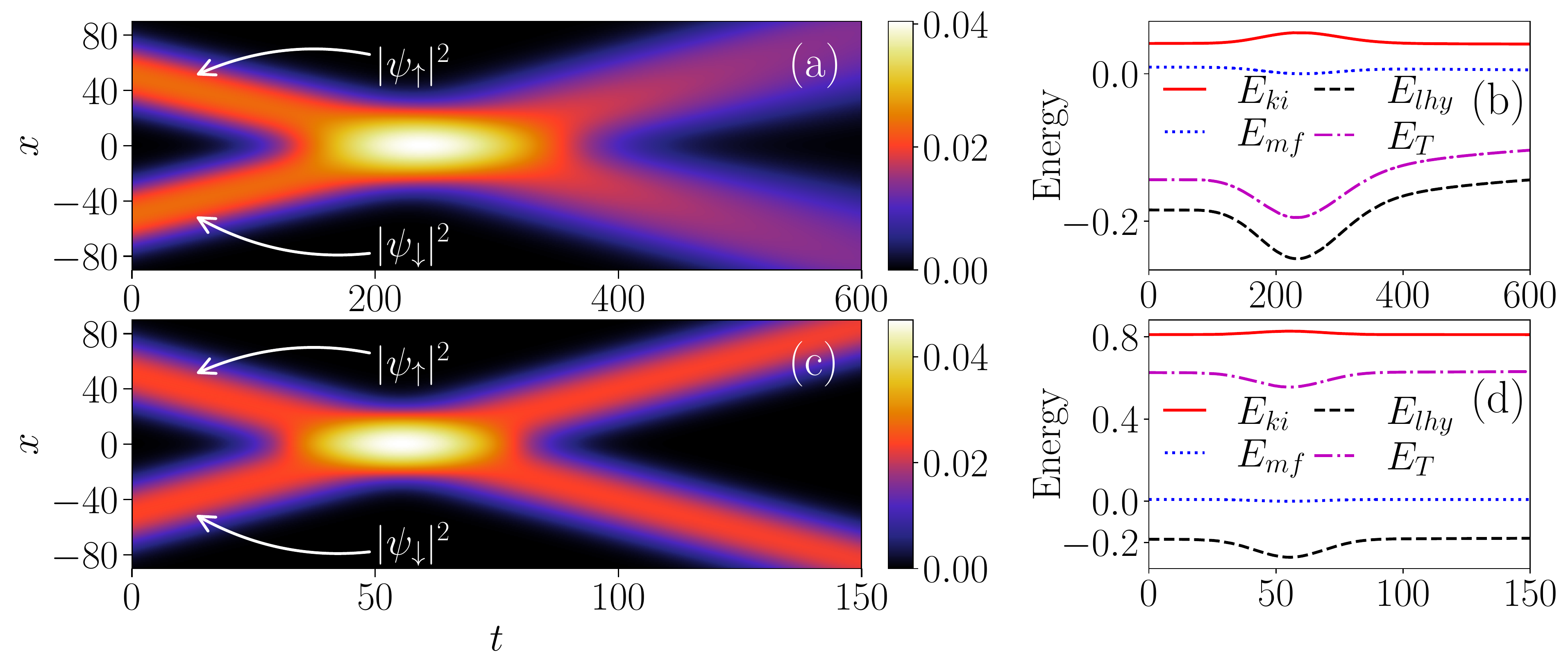}
\caption{(a) The total density plots depicting the inelastic collision dynamics between the quantum soliton for $g =- g_{\uparrow\downarrow} = 0.5$ and $\Omega = k_L = 0$ as the velocity given to the up and down component is in the direction of $-x$ and $x$ direction respectively, with magnitude $v = 0.2$.  (b) Evolution of the different energies, $E_{ki}$, $E_{lhy}$, $E_{mf}$ and total energy ($E_T$) with time for collision as shown in (a). There is a significant increase in the $E_{lhy}$ and total energy ($E_T$) after the collision, indicating the inelastic nature of the same. (c) plots of the total density depicting the elastic collision dynamics of quantum soliton for $v = 0.9$. (d) Evolution of the different energies with time for the collision as shown in (c). Before and after the collision, all the energy remains the same, indicating the elastic collision nature.}
\label{fig:bi-col}
\end{figure*}
In Fig.~\ref{fig:quench2b_case4} we show the dynamics that solely arises due to the quenching of the Rabi coupling  for the ground states initially prepared for $g = - g_{\uparrow\downarrow} = 1.0$ and  $\Omega = k_L = 2$. We quench the parameters as (a) $\Omega = 2 \to 1$ and in (b) $\Omega = 2 \to 3$ at $t=10$.  Here we notice that the stripe soliton exhibits expansion upon the quenching of the parameters. We find that the breathing oscillation frequency increases on increasing the interaction strength, as shown in Fig.~\ref{fig:breath1}(c). Further we  consider another stripe soliton with $\Omega = k_L = 2$ in which the Rabi coupling  is quenched as $\Omega = 2 \to 1$. For this, we find that the frequency of the striped soliton breather becomes $\omega_{bf} \sim 0.08$ upon increasing the interaction strength. When we change $\Omega = 2\to 3$, we obtain that the initial profile, which has two maximum peaks and two sidelobes, gets transformed into the multiple side lobes after the quench. Apart from this, the innermost soliton appears to repel each other after a finite time as they approach each other. A similar kind of dynamical behaviour gets repeated over time.

\begin{figure*}[!ht]
\centering\includegraphics[width=0.99\linewidth]{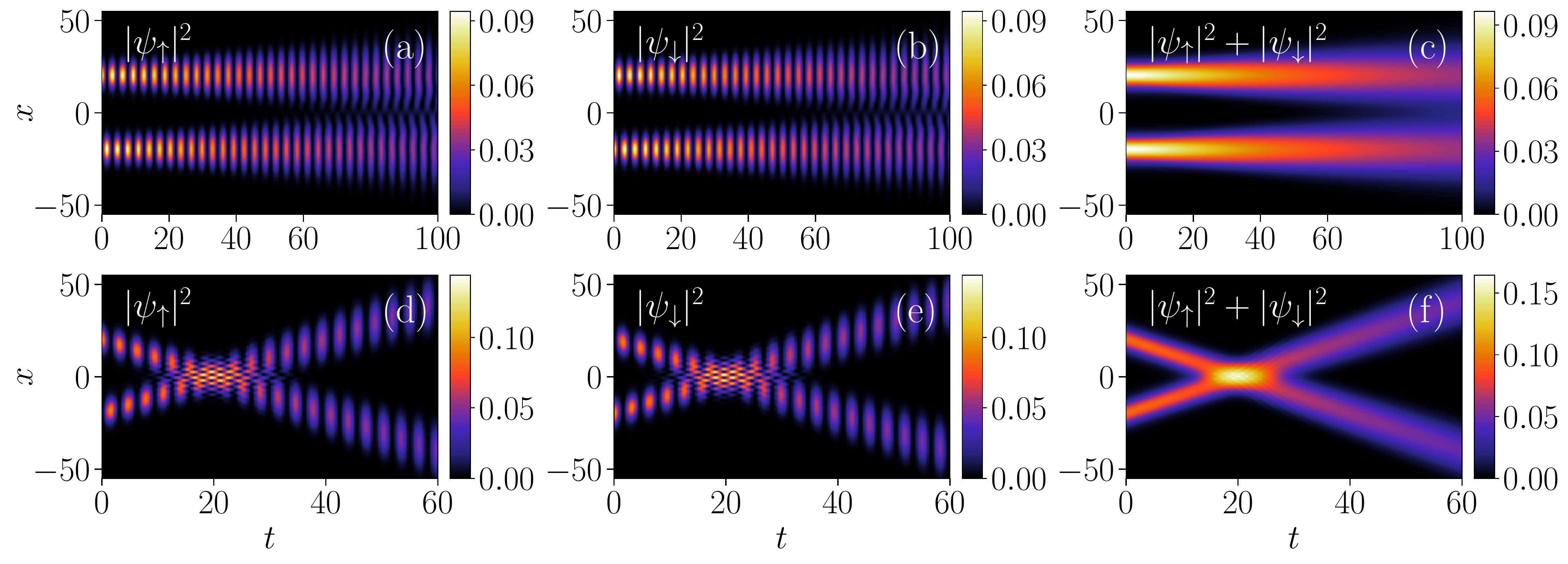}
\caption{Collisional dynamics of the quantum solitons prepared with $g=-g_{\uparrow\downarrow}=0.5$ and $\Omega=1$, $k_L=0$ and at $t=0$ with velocity given to the up and down component with opposite value and magnitude as (a)-(c): $v=0$ and (d)-(f): $v=1$. For $v=0$, the soliton displays spatio-temporal breathers with propagation along a line. However, for $v=1$ soliton exhibits space-time breathers with quasi-elastic collision around $t\sim 20$.}
\label{fig:collision-quench1}
\end{figure*}

\begin{figure*}[!ht]
\centering\includegraphics[width=0.99\linewidth]{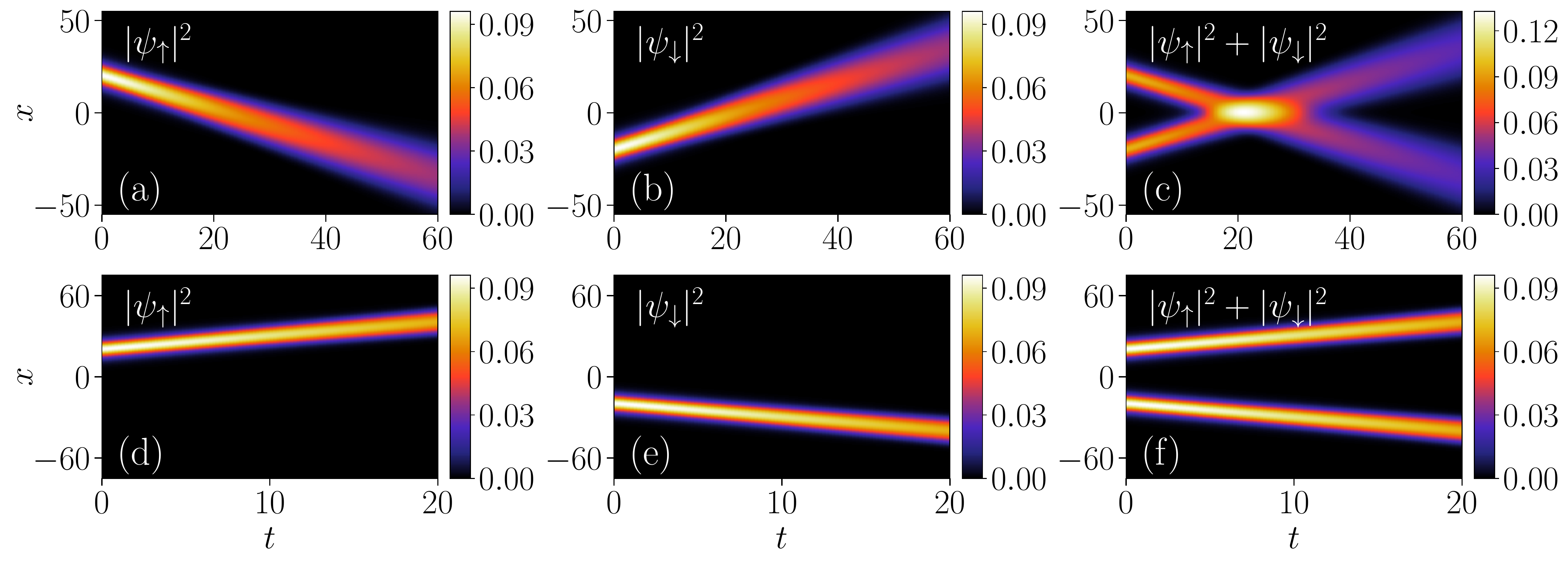}
\caption{Collisional dynamics of the quantum solitons prepared with $g=-g_{\uparrow\downarrow}=0.5$ $\Omega=0$ and $k_L=0.5$ and the individual components are given equal and opposite velocity ($v=0.5$) at $t=0$. Also quenching on SO coupling parameter is performed as (a)-(c): $k_L=0.5 \to 0.1$  and (d)-(f): $k_L=0.5 \to 2$. Upon quenching the SO coupling, the solitons undergo inelastic collision in (a)-(c), while  solitons repel each other for (d)-(f).}
\label{fig:collision-quench2}
\end{figure*}
\begin{figure*}[!ht]
\centering\includegraphics[width=0.99\linewidth]{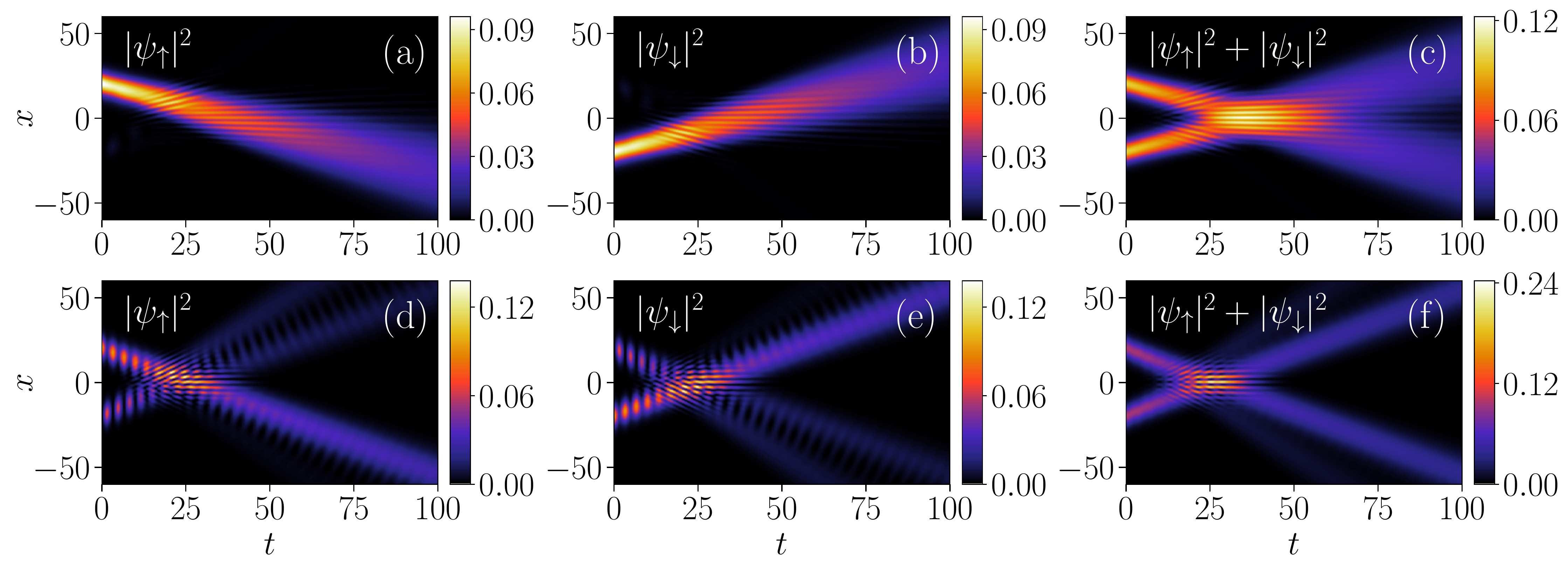}
\caption{Collisional dynamics of the quantum solitons prepared with the same parameters and given same velocity at $t=0$ as those for Fig.~\ref{fig:collision-quench2}. At $t=0$ the Rabi coupling is quenched as (a)-(c): $\Omega=0 \to 0.1$ (d)-(f): $\Omega=0 \to 0.8$. Upon quenching, Rabi coupling quantum soliton undergoes inelastic collision and displays an interference pattern (a)-(c). However, the soliton displays quasi-elastic collision with secondary solitons generation in (d)-(f).}
\label{fig:collision-quench3}
\end{figure*}
\subsection{Collisional dynamics  of quantum soliton}
\label{sec:4e}

After studying the quenching dynamics of the quantum soliton by perturbing it with an initial velocity, in this section, we investigate how the solitons undergo different kinds of collisions that affect its overall stability depending upon the magnitude of the initial velocity. In particular, we find elastic, inelastic, repulsive and space-time breather soliton against collisions.

In absence of the trap, the quantum solitons are set in motion by giving an initial velocity $\mp v$ to the stationary ground state $(\psi_1,\psi_2)$ with a multiplier $\exp{(\mp ivx)}$ respectively~\cite{Astrakharchik2018}. Here we start for the case when the solitons are positioned at $\pm 50$  for $g=-g_{ab}=0.5$ and $\Omega=k_L=0$. For a weak velocity ($v=0.2$) the solitons undergo inelastic collision as depicted in Fig.~\ref{fig:bi-col}(a).  We find that the densities before the collision and after the collision are not the same. This indicates that after the collision, the solitons are unable to retain their shape and size same as those before the collision. As the initial positions of the soliton are at $\pm 50$ the expected time for collision is $t=x/v=250$ which can be clearly seen in Fig.~\ref{fig:bi-col}(a).  We find that after the collision ($t\sim 400$), the solitons start exhibiting the expansion, which makes them quite different in shape and size than they were before the collision. In order to understand the inelastic collision in better way in Fig.~\ref{fig:bi-col}(b) we plot the time evolution of the different energy contribution, like, kinetic energy ($E_{ki}$), mean-field energy ($E_{mf}$), LHY energy ($E_{lhy}$) and the total energy ($E_T$).  The explicit form of energies is provided in the Eq.~(\ref{eq:energy}) of Appendix~\ref{app:ene}. We find that the mean-field energy, kinetic, LHY and total energy after the collision does not remain same as they are before the collisions which is consistent with the inelastic nature of the collision observed in the classical case. 

For large velocity ($v=0.9$), we observe the elastic collision with no change in the shape of the soliton after the collision. In  Fig.~\ref{fig:bi-col} (c) we show the total density corresponding to the soliton when each component has been given an initial equal velocity ($v=0.9$) but opposite direction with their initial positions  at $\pm 50$ for $g=-g_{ab}=0.5$ and $\Omega=k_L=0$. For this, the expected collision time will be at $t=55.5$. We find that the densities before and after the collision are almost similar. In Fig. ~\ref{fig:bi-col} (d), we plot the time evolution of the different energy contributions to confirm the elastic nature of the collision. It is observed that the mean-field energy, kinetic, LHY and total energy remain the same after the collisions what they were before the collision, which are consistent with the elastic nature of the collision. This behaviour is consistent with the experimental observations of the collision between the quantum droplet in three dimension~\cite{Ferioli2019}.

So far, we have analyzed the collisional dynamics of the binary BECs with LHY correction and in the absence of Rabi and SO coupling parameters. However, as we consider it for finite SO and Rabi couplings, we find that the solitons get degenerated and thus, it could not provide the relevant collision dynamics. Considering what follows, we present the collision dynamics by keeping either SO or Rabi coupling frequency to be finite. Here as we obtain the spatially separated soliton, we employ the quenching of velocity and the coupling parameter, which was zero while preparing the states. To study the collisional dynamics with finite coupling parameters, we prepare the initial ground state with $\Omega=1$ and $k_L=0$. In Fig.~\ref{fig:collision-quench1}(a)-(c) we show the dynamics with zero velocity. For this case, we find each spin component degenerates in the $\pm x$ direction and further exhibits the space-time breathers, which experience expansion after the finite time ($t\sim 40$). As we pass on the finite velocity, the solitons start interacting with each other. As for an example, for $v=1$ the soliton starts moving towards each other that results in the elastic collision  between them at $t\sim 20$ [see Fig.~\ref{fig:collision-quench1}(d)-(f)]. Note that here, after the collision, there is a small change in the density, which is different from the collision behaviour observed in Fig.~\ref{fig:bi-col}(c). Therefore, we refer to this as the quasi-elastic collision.

Next, we present the collisional dynamics by preparing the ground states with finite SO coupling and zero Rabi coupling frequency. We consider the system with $\Omega=0$, $k_L=0.5$ with $g=0.5$ and finite velocity $v=0.5$. Once we prepare the ground state, we analyse the collisional dynamics by quenching the SO coupling for two different situations. When $v > k_L/2$ non-degenerated soliton with  decaying  or expansion feature undergo inelastic collision as depicted in Fig.~\ref{fig:collision-quench2}(a)-(c). However, for $v < k_L/2$ we observe that the soliton appears to repel each other upon progression of time [see Fig.~\ref{fig:collision-quench2}(d)-(f)]. 

Further, we analyse the collision dynamics by quenching the Rabi coupling. Like in the previous case, we study the dynamics of the solitons for two different situations. When $\Omega < k_L^2$ non-degenerated soliton undergo inelastic collision. They appear to show the stripe fringes as depicted in Fig.~\ref{fig:collision-quench3}(a)-(c). For $\Omega > k_L^2$, we find a collision of bright soliton is quasi-elastic with the generation of interference stripe fringes. We also witness the generation of secondary solitons, as shown in Fig.~\ref{fig:collision-quench3}(d)-(f).

\section{Summary and Conclusions}
\label{sec:5}
Using the mean-field model along with the LHY correction, we have numerically investigated the structure and dynamics of the ground state of the self-bound state in 1D spin-orbit coupled BECs. Depending upon the nature of Rabi and SO couplings, the ground state is either quantum bright or stripe soliton. We deduced an analytical solution of the quantum bright soliton without Rabi coupling. We further studied the dynamics of the soliton using the three protocols, namely, by giving the initial velocity to the soliton, quenching the SO and Rabi coupling parameters and allowing the component to collide by attributing an equal and opposite initial velocity. We found that the velocity perturbations generate breathing-like soliton. The breathing frequency increases upon the increase in the SO coupling for a given Rabi coupling frequency attains a maximum at the SO coupling, where the phase transition from the quantum bright to stripe soliton takes place. The magnitude of breathing frequency at critical SO coupling remains independent of the initial velocity and exhibits a power law dependence on the Rabi coupling frequency with an exponent $\sim 0.16$. We found that the critical breathing frequency increases upon increasing the interaction strengths. 
Using the contribution of the kinetic, SO, and Rabi energy terms, we find that while the domination of the breathing soliton phase by the SO and Rabi coupling, the moving soliton phase is the result of the kinetic energy part. This particular feature leads the attainment of the minima of the chemical potential at the critical velocity where change in the phase takes place. 
We have realized the presence of several dynamical phase transitions, like, QBS to QSS and the multisoliton behaviour depending upon the quenching protocol of the Rabi coupling parameters. By quenching both Rabi and the SO couplings, we can control the direction/angle of the inner- and outer soliton. The quenching of Rabi and SO couplings facilitate several interesting dynamical phases like a repulsive soliton, space-time breathers, filamentation, etc.

We have also analyzed the collision dynamics of solitons. Depending upon the velocity of the soliton, we observed the presence of elastic and inelastic collisions. An inelastic collision occurs for low velocities, while it exhibits elastic collision at high velocities consistent with earlier experimental observations~\cite{Ferioli2019}. We have also complemented the collision dynamics by analyzing the nature of different energy terms like kinetic energy, mean-field energy, and LHY. Also, we analyzed the collision dynamics of the solitons by quenching either the velocity or the SO/Rabi coupling. For the former, we found the presence of quasi-elastic collision, while for the latter, we realize inelastic collision.


In this paper, we have considered the dynamical evolution of the quantum solitons for the situation when the mean-field contribution is negligible. It would be interesting to explore the dynamics in the similar line presented in this work in the presence of significant mean-field contribution along with the LHY correction in the quantum soliton.

\acknowledgments 
We thank Cesar Cabrera for his comments on the initial draft of our manuscript. S.G. would like to acknowledge the financial support from the University Grants Commission - Council of Scientific and Industrial Research (UGC-CSIR), India. R.R. acknowledge support from the Ministry of Science and Technology (MOST), Taiwan, under Grant No. MOST-111-2119-M-001-002. P.K.M. acknowledges the Department of Science and Technology - Science and Engineering Research Board (DST-SERB) India for the financial support through  Project No. ECR/2017/002639. The work of P.M. is supported by DST-SERB under Grant No. CRG/2019/004059, DST-FIST under Grant No. SR/FST/PSI-204/2015(C), and MoE RUSA 2.0 (Physical Sciences). 

\onecolumngrid
\appendix

\section{Calculation of energy in the SO coupled BECs with LHY correction}\label{app:ene}
In this appendix, we provide the detailed steps to obtain the total energy of the SO coupled BECs with LHY correction. The stationary state solution is given by 
\begin{align} \label{chem}
\psi_{j}(x,y) = \big(\psi_{jR} + \mathrm{i} \psi_{jI} \big) e^{-\mathrm{i} \mu_{j} t}
\end{align}
where $j\in \{\uparrow, \downarrow\}$, $\psi_{jR}$ and $ \psi_{jI}$ are the real and imaginary parts of the stationary wavefunction respectively,  and $\mu_{\uparrow, \downarrow}$ are the chemical potential of the spin-up and down components respectively~\cite{Ravisankar2021a}. As we insert this solution in the Eqs.~(\ref{eq:gpsoc:1}) we get
\begin{subequations}
\label{eq:gpsoc:reim}
\begin{align}
\mu_{\uparrow} \psi_{\uparrow R} = & \left[ -\frac{1}{2}\frac{\partial^2 }{\partial x^2} + g \lvert \psi_\uparrow \rvert^2 
+ g_{\uparrow\downarrow} \lvert \psi_\downarrow \rvert^2 - \frac{g^{3/2}}{\pi} \sqrt{\lvert \psi_{\uparrow}\rvert^2+\lvert \psi_{\downarrow}\rvert^2}\right] \psi _{\uparrow  R} 
+ k_{L} \left(\frac{\partial \psi_{\uparrow I}}{\partial x} \right) + \Omega  \psi_{\downarrow R}, \label{eq:gpsoc2:reim1} \\
\mu_{\downarrow} \psi_{\downarrow R}  = & \left[ -\frac{1}{2}\frac{\partial^2 }{\partial x^2}  + g \lvert \psi_\downarrow \rvert^2 
+ g_{\downarrow\uparrow} \lvert \psi_\uparrow \rvert^2 - \frac{g^{3/2}}{\pi} \sqrt{\lvert \psi_{\uparrow}\rvert^2+\lvert \psi_{\downarrow}\rvert^2}\right]\psi _{\downarrow R}
- k_{L}\left(\frac{\partial \psi_{\downarrow I}}{\partial x}\right) + \Omega \psi_{\uparrow R}, \label{eq:gpsoc2:reim2} 
\end{align}
and for the imaginary parts we have
\begin{align}
\mu_{\uparrow} \psi_{\uparrow I} = & \left[-\frac{1}{2}\frac{\partial^2 }{\partial x^2}  + g \lvert \psi_\uparrow \rvert^2 
+ g_{\uparrow\downarrow} \lvert \psi_\downarrow \rvert^2 - \frac{g^{3/2}}{\pi} \sqrt{\lvert \psi_{\uparrow}\rvert^2+\lvert \psi_{\downarrow}\rvert^2}\right] \psi _{\uparrow  I} 
- k_{L} \left(\frac{\partial \psi_{\uparrow R}}{\partial x} \right) + \Omega  \psi_{\downarrow I} , \label{eq:gpsoc2:reim3}\\
\mu_{\downarrow} \psi_{\downarrow I}  = & \left[ -\frac{1}{2}\frac{\partial^2 }{\partial x^2}  + g \lvert \psi_\downarrow \rvert^2 
+ g_{\downarrow\uparrow} \lvert \psi_\downarrow \rvert^2 - \frac{g^{3/2}}{\pi} \sqrt{\lvert \psi_{\uparrow}\rvert^2+\lvert \psi_{\downarrow}\rvert^2}\right] \psi _{\uparrow  I} 
+ k_{L} \left(\frac{\partial \psi_{\downarrow R}}{\partial x} \right) + \Omega  \psi_{\uparrow I}, \label{eq:gpsoc2:reim4}
\end{align}
\end{subequations}
where $ \lvert \psi_\uparrow \rvert^2  =   \psi_{\uparrow R}^2 +  \psi_{\uparrow I}^2  $ and  $ \lvert \psi_\downarrow \rvert^2  =   \psi_{\downarrow R}^2 +  \psi_{\downarrow I}^2  $.
Multiplying Eq.~(\ref{eq:gpsoc2:reim1}) with $\psi_{\uparrow R}$ and Eq.~(\ref{eq:gpsoc2:reim2}) with $\psi_{\downarrow R}$, and integrating we get
\begin{subequations}
\begin{align}
\mu_{\uparrow} \int dx \, \psi_{\uparrow R}^2 = & \int dx\, \psi _{\uparrow  R}\left\{\left[  -\frac{1}{2}\frac{\partial^2 }{\partial x^2}  + g \lvert \psi_\uparrow \rvert^2 
+ g_{\uparrow\downarrow} \lvert \psi_\downarrow \rvert^2 - \frac{g^{3/2}}{\pi} \sqrt{\lvert \psi_{\uparrow}\rvert^2+\lvert \psi_{\downarrow}\rvert^2}\right]\psi _{\uparrow  R} 
+   k_{L} \left(\frac{\partial \psi_{\uparrow I}}{\partial x} \right)
+ \Omega \psi_{\downarrow R}   \right\}, \label{eq:gpsoc2:mu1} \\
\mu_{\downarrow} \int dx \,\psi_{\downarrow R}^2 = & \int dx\, \psi _{\downarrow  R} \left\{ \left[-\frac{1}{2}\frac{\partial^2 }{\partial x^2}  + g \lvert \psi_\downarrow \rvert^2 
+ g_{\downarrow\uparrow} \lvert \psi_\uparrow \rvert^2 - \frac{g^{3/2}}{\pi} \sqrt{\lvert \psi_{\uparrow}\rvert^2+\lvert \psi_{\downarrow}\rvert^2} \right] \psi _{\downarrow  R} 
-   k_{L} \left(\frac{\partial \psi_{\downarrow I}}{\partial x} \right)
+ \Omega  \psi_{\uparrow R}   \right\}  , \label{eq:gpsoc2:mu2}
\end{align}
\end{subequations}
Similarly, from the imaginary part equations (Eqs.~\ref{eq:gpsoc2:reim3} and \ref{eq:gpsoc2:reim4}) we obtain
\begin{subequations}
\begin{align}
\mu_{\uparrow} \int dx \, \psi_{\uparrow I}^2 = & \int dx\, \psi _{\uparrow  I}\left\{\left[  -\frac{1}{2}\frac{\partial^2 }{\partial x^2}  + g \lvert \psi_\uparrow \rvert^2 
+ g_{\uparrow\downarrow} \lvert \psi_\downarrow \rvert^2 - \frac{g^{3/2}}{\pi} \sqrt{\lvert \psi_{\uparrow}\rvert^2+\lvert \psi_{\downarrow}\rvert^2}\right]\psi _{\uparrow  I} 
-  k_{L} \left(\frac{\partial \psi_{\uparrow R}}{\partial x} \right)
+ \Omega \psi_{\downarrow I}   \right\}, \label{eq:gpsoc2:mu1} \\
\mu_{\downarrow} \int dx \,\psi_{\downarrow I}^2 = & \int dx\, \psi _{\downarrow  I} \left\{ \left[-\frac{1}{2}\frac{\partial^2 }{\partial x^2}  + g \lvert \psi_\downarrow \rvert^2 
+ g_{\downarrow\uparrow} \lvert \psi_\uparrow \rvert^2 - \frac{g^{3/2}}{\pi} \sqrt{\lvert \psi_{\uparrow}\rvert^2+\lvert \psi_{\downarrow}\rvert^2} \right] \psi _{\downarrow  I} 
+  k_{L} \left(\frac{\partial \psi_{\downarrow R}}{\partial x} \right)
+ \Omega \psi_{\uparrow I} \right\}, \label{eq:gpsoc2:mu21}
\end{align}
\end{subequations}

Upon rearranging the above equations (Eqs.~\ref{eq:gpsoc2:mu2}-\ref{eq:gpsoc2:mu21}) we obtain
\begin{subequations}
\label{eq:gpsoc2:mu:ab}
\begin{align}
\mu_{\uparrow} = & \frac{1}{ \int \lvert\psi_{\uparrow}\rvert^2dx} \int \left[ \frac{1}{2} \left|\frac{\partial \psi_{\uparrow }}{\partial x}\right|^2 + \left\{g \lvert \psi_\uparrow \rvert^2 
+ g_{\uparrow\downarrow} \lvert \psi_\downarrow \rvert^2 - \frac{g^{3/2}}{\pi} \sqrt{\lvert \psi_{\uparrow}\rvert^2+\lvert \psi_{\downarrow}\rvert^2} \right\}\lvert\psi_{\uparrow }\rvert^2\right] dx\notag \\
& + \frac{1}{ \int\vert\psi_{\uparrow }\rvert^2 dx } \int \left[ k_{L} \left(\frac{\partial \psi_{\uparrow I}}{\partial x} \right)
+ \Omega \psi_{\downarrow R} \right] \psi_{\uparrow R} dx + \frac{1}{ \int\vert\psi_{\uparrow }\rvert^2 dx } \int \left[ -k_{L} \left(\frac{\partial \psi_{\uparrow R}}{\partial x}\right)
+ \Omega \psi_{\downarrow I} \right] \psi_{\uparrow I} dx, \label{eq:gpsoc2:mu:1} \\
\mu_{\downarrow} = & \frac{1}{ \int \lvert\psi_{\downarrow }\rvert^2dx} \int \left[ \frac{1}{2} \left|\frac{\partial \psi_{\downarrow }}{\partial x}\right|^2 +\left\{ g \lvert \psi_\downarrow \rvert^2 
+ g_{\downarrow\uparrow} \lvert \psi_\downarrow \rvert^2 - \frac{g^{3/2}}{\pi} \sqrt{\lvert \psi_{\uparrow}\rvert^2+\lvert \psi_{\downarrow}\rvert^2} \right\}\lvert\psi_{\downarrow }\rvert^2\right] dx \notag \\
& + \frac{1}{ \int\vert\psi_{\downarrow }\rvert^2 dx} \int  \left[ -k_{L} \left(\frac{\partial \psi_{\downarrow I}}{\partial x} \right)
+ \Omega \psi_{\uparrow R} \right] \psi_{\downarrow R} dx+ \frac{1}{ \int\lvert\psi_{\downarrow }\rvert^2 dx} \int \left[  k_{L} \left(\frac{\partial \psi_{\downarrow R}}{\partial x}  \right)
+ \Omega \psi_{\uparrow I} \right] \psi_{\downarrow I} dx, \label{eq:gpsoc2:mu:2}
\end{align}
\end{subequations}
From the above equations we obtain the different contribution in the energy as
\begin{subequations}
\label{eq:energy}
\begin{align}
E_{ki}= & \int \left[\frac{1}{ 2\int \lvert\psi_{\uparrow }\rvert^2dx} \left|\frac{\partial \psi_{\uparrow }}{\partial x}\right|^2 + \frac{1}{ 2\int \lvert\psi_{\downarrow }\rvert^2dx} \left|\frac{\partial \psi_{\downarrow }}{\partial x}\right|^2 \right]dx \\   
E_{mf}= &\int\left[\frac{1}{ \int \lvert\psi_{\uparrow }\rvert^2dx}\left[\frac{g}{2} \lvert \psi_{\uparrow} \rvert^2 + \frac{g_{\uparrow\downarrow}}{2} \lvert \psi_{\downarrow} \rvert^2\right]\lvert\psi_{\uparrow }\rvert^2 +\frac{1}{ \int \lvert\psi_{\downarrow }\rvert^2 dx} \left[\frac{g_{\downarrow\uparrow}}{2} \lvert \psi_{\uparrow} \rvert^2+\frac{g}{2} \lvert \psi_{\downarrow} \rvert^2\right]\vert\psi_{\downarrow }\rvert^2\right]dx \\
E_{lhy} = & \int\left[-\frac{1}{ \int \lvert\psi_{\uparrow }\rvert^2 dx}\left[\frac{2}{3\pi}g^{3/2} \sqrt{\lvert \psi_{\uparrow} \rvert^2+\lvert \psi_{\downarrow} \rvert^2}\right] \lvert\psi_{\uparrow }\rvert^2 - \frac{1}{ \int \lvert\psi_{\downarrow }\rvert^2dx}\left[\frac{2}{3\pi}g^{3/2} \sqrt{\lvert \psi_{\uparrow} \rvert^2+\lvert \psi_{\downarrow} \rvert^2}\right] \lvert\psi_{\downarrow }\rvert^2\right]dx \\
E_{SO} = & \int \left\{\frac{1}{ \int \lvert\psi_{\uparrow }\rvert^2 dx}\left[ \left(k_L\frac{\partial \psi_{\uparrow I}}{\partial_x}  \right)\psi_{\uparrow R}+\left(-k_L\frac{\partial \psi_{\uparrow R}}{\partial_x} \right)\psi_{\uparrow I}\right]
+\frac{1}{ \int \lvert\psi_{\downarrow }\rvert^2 dx} \left[\left(-k_L\frac{\partial \psi_{\downarrow I}}{\partial_x}  \right)\psi_{\downarrow R}+\left(k_L\frac{\partial \psi_{\downarrow R}}{\partial_x}  \right)\psi_{\downarrow I}\right]\right\}dx \\
E_{Rabi} = & \int\left\{\frac{1}{ \int \lvert\psi_{\uparrow }\rvert^2 dx}\left[\left(\Omega \psi_{\downarrow R}  \right)\psi_{\uparrow R} +\left(\Omega \psi_{\downarrow I}  \right)\psi_{\uparrow I}\right]
+\frac{1}{ \int \lvert\psi_{\downarrow }\rvert^2 dx}\left[ \left(\Omega \psi_{\uparrow R} \right)\psi_{\downarrow R}
+\left(\Omega \psi_{\uparrow I} \right)\psi_{\downarrow I}\right]\right\}dx
\end{align}
\end{subequations}
where $E_{ki}$ represents the kinetic energy, $E_{mf}$ mean-field contribution, $E_{lhy}$ is the energy due to the LHY correction, and $E_{SO}$ is the contribution due to the SO coupling. The total energy $E_T=E_{ki}+E_{mf}+E_{lhy}+E_{SO}$.

\twocolumngrid
%

\end{document}